%% file: main.tex
\documentclass[aps,prl,twocolumn,showpacs,superscriptaddress]{revtex4-2}

\usepackage{amsmath}
\usepackage{graphicx}
\usepackage{lmodern}
\usepackage{multirow}

\usepackage{amssymb}
\usepackage{tikz-feynman}
\usepackage{bm}
\usepackage{braket}
\usepackage{mathtools}

\newcommand{\mr}{moir\'{e} }

\DeclareMathOperator{\Tr}{Tr}

\usepackage[dvipsnames]{xcolor}
\usepackage[colorlinks=true]{hyperref}
\hypersetup{
    colorlinks=true,
    linkcolor=blue,
    citecolor=blue,
    urlcolor=blue
} 

\newcommand{\cre}[2]{\hat{#1}^\dagger_{#2}}
\newcommand{\des}[2]{\hat{#1}_{#2}}

\def\kk{\mathbf{k}}
\def\qq{\mathbf{q}}
\def\pp{\mathbf{p}}

\def\GG{\mathbf{G}}
\def\QQ{\mathbf{Q}}

\makeatletter
\def\maketitle{
\@author@finish
\title@column\titleblock@produce
\suppressfloats[t]}
\makeatother

\makeatletter
\newcommand{\appendixlocaltoc}{%
\begingroup
\let\addcontentsline\oldaddcontentsline
\let\clearpage\relax
\section*{Appendix Contents}
\endgroup
\@starttoc{atc}%
}
\newcommand{\startappendixlocaltoc}{%
\addtocontents{atc}{\protect\setcounter{tocdepth}{3}}
\global\let\oldaddcontentsline\addcontentsline
\renewcommand{\addcontentsline}[3]{%
\ifstrequal{##1}{toc}{%
\ifstrequal{##2}{section}{\oldaddcontentsline{atc}{##2}{##3}}{%
\ifstrequal{##2}{subsection}{\oldaddcontentsline{atc}{##2}{##3}}{%
\ifstrequal{##2}{subsubsection}{\oldaddcontentsline{atc}{##2}{##3}}{}%
}%
}%
}{%
\oldaddcontentsline{##1}{##2}{##3}%
}%
}%
}
\makeatother

\begin{document}

\title {Weak-Coupling Pair-Density-Wave from Momentum-Space Nonsymmorphic Symmetry}
\author{Ming-Rui Li}
\thanks{These authors contributed equally to the work.}
\affiliation{Chen-Ning Yang Institute for Advanced Study (YIAS), Tsinghua University, Beijing 100084, China}
\author{Zhengzhi Wu}
\thanks{These authors contributed equally to the work.}
\affiliation{Rudolf Peierls Centre for Theoretical Physics, Parks Road, Oxford, OX1 3PU, UK}
\author{Hong Yao}
\email{yaohong@tsinghua.edu.cn}
\affiliation{Chen-Ning Yang Institute for Advanced Study (YIAS), Tsinghua University, Beijing 100084, China}
\date{September 22, 2026}

\begin{abstract}
Pair-density-wave (PDW) order is a superconducting state with a spatially modulated order parameter and is generally subleading to uniform superconductivity within the conventional weak-coupling BCS paradigm. Here, using a twisted bilayer checkerboard-lattice model, we show that momentum-space nonsymmorphic symmetry provides a generic mechanism that can instead promote PDW order to a leading weak-coupling instability. Combining a controlled Wilsonian renormalization-group analysis at charge neutrality with a finite-doping Bethe--Salpeter analysis, we determine the leading ordering tendencies among various competing instabilities. At charge neutrality, two symmetry-related quadratic band touchings support  spin-singlet and spin-triplet PDW instabilities for different attractive interactions, whereas repulsive interactions favor  quantum Hall insulators and finite-$\mathbf Q$ density-wave orders. Upon doping, the nonsymmorphic symmetry, together with time reversal, protects the Cooper logarithm in the finite-momentum $\mathbf Q$ pairing channel at generic fillings, allowing PDW order to remain a leading instability over the broad doping range studied. This mechanism extends naturally to \mr Chern bands: by breaking time-reversal symmetry while preserving the nonsymmorphic symmetry and inversion, we obtain a fully gapped topological spin-triplet PDW state with BdG Chern number $C_{\text{BdG}}=8$. Our results establish momentum-space nonsymmorphic symmetry as a general weak-coupling route to PDW superconductivity, including topological PDW states.
\end{abstract}

\maketitle

\textit{Introduction.---}
Pair-density waves (PDWs) are superconducting states with spatially modulated order parameters and, ideally, no uniform component~\cite{annurev}. Unlike the Fulde-Ferrell-Larkin-Ovchinnikov state~\cite{PhysRev.135.A550,LO,RevModPhys.76.263}, a PDW can arise without an externally imposed spin imbalance or magnetic field. By combining superconducting coherence with broken translation symmetry, it induces secondary charge order and can support vestigial phases
such as charge-$4e$ superconductivity, fractional vortices, paired fractional quantum Hall fluids, and unconventional quantum criticality~\cite{PhysRevB.79.064515,RevModPhys.87.457,Berg_2009,Berg2009,PhysRevB.89.165126,PhysRevX.9.021047,annurev}. Accumulating experimental evidence for PDW states has been reported in a variety of material platforms~~\cite{Hamidian2016,Ruan2018,Edkins2019,Du2020,PhysRevX.11.011007,Liu2021TMDPDW,Chen2021,Deng2024KagomePDW,Han2025KagomePDW,Gu2023,Zhao2023,Liu2023,Kong2025IronPDM,Wang2026MoireCPDM,PhysRevB.83.104506,PhysRevB.85.134513,Shi2020,PhysRevLett.99.127003,Aishwarya2023}, while PDW order has also attracted extensive theoretical interest~\cite{PhysRevLett.88.117001,PhysRevB.76.140505,Yang_2009,PhysRevB.82.041102,PhysRevLett.105.146403,PhysRevLett.107.187001,PhysRevB.85.035104,PhysRevLett.113.046402,PhysRevLett.114.197001,PhysRevB.95.155116,PhysRevLett.122.167001,PhysRevLett.125.167001,Huang2022HolsteinHubbardPDW,PhysRevB.112.L140505,PhysRevX.4.031017,PhysRevB.102.235423,doi:10.1126/sciadv.aat4698,Setty2023,PhysRevB.108.035135,PhysRevLett.130.126001,PhysRevLett.131.016001,PhysRevB.108.L201110,PhysRevB.110.094515,PhysRevLett.133.176501,PhysRevLett.130.026001,PhysRevB.107.045122,PhysRevLett.131.016002,PhysRevLett.131.026601,PhysRevX.14.041004,Barlas2025Kekule,wu2026exactlysolvablepairdensitywave,wu2026exactpairdensitywave,durrnagel2026pairdensitywaveorder}.

Nevertheless, a generic mechanism for PDW order at weak coupling remains elusive. In the weak-coupling limit, a PDW instability requires perfect nesting of the Fermi surface in a finite-momentum particle-particle channel:
$
\xi(\mathbf{k})=\xi(\mathbf{Q}-\mathbf{k})$, so that the corresponding Cooper susceptibility acquires the same logarithmic divergence characteristic as conventional BCS pairing. Such a nesting condition is not generically satisfied. Previous proposals have shown that it can arise under more restrictive circumstances, for example in the presence of  external fields tuned to special values \cite{PhysRevLett.130.126001,PhysRevB.108.035135}, near special fillings such as van Hove singularities \cite{PhysRevLett.131.026601}, or at sufficiently low carrier density where the dispersion can be approximated as parabolic \cite{Hsu2017}. A general mechanism that enforces finite-momentum particle-particle nesting at generic fillings without any external fields, however, remains lacking.

Here we show that particle-particle nesting can be enforced by a momentum-space nonsymmorphic symmetry naturally realized in a twisted bilayer checkerboard (TBCB) lattice~\cite{PhysRevResearch.4.043151,PhysRevB.110.165142,wu2026exactpairdensitywave}. Moir\'e systems provide tunable settings for such symmetry-engineered band structures and correlated phases~\cite{BistritzerMacDonald2011,Cao2018Superconductivity,Cao2018Correlated,Andrei2021Marvels,Kennes2021Simulator,PhysRevLett.122.086402,PhysRevLett.123.237002,PhysRevB.101.060505,PhysRevLett.124.167002,PhysRevLett.129.047601,PhysRevB.103.205415,PhysRevLett.132.096602,Calugaru2025,5zt2-scbg,htld-vgws,klebl2026extendedswavesuperconductivitympoint,xu2026organizingprinciplesmoirequantum}. Away from its magic-angle flat-band limits, the TBCB model hosts two $C_{4z}$- and time-reversal-protected quadratic band touchings (QBTs) at $\Gamma$ and $M$, separated by $\mathbf M=(\pi,\pi)$~\cite{PhysRevResearch.4.043151}. Their finite density of states makes short-range interactions marginal at tree level~\cite{PhysRevLett.103.046811,PhysRevB.82.205106}. Unlike the QBTs in Bernal-stacked bilayer graphene, which are split by symmetry-allowed trigonal warping into Dirac cones which remain stable below a finite interaction threshold~\cite{PhysRevB.82.205106,PhysRevB.79.085116,PhysRevB.90.035126,PhysRevLett.114.237001,PhysRevLett.118.166802}, the TBCB QBTs are symmetry protected and require no additional fine tuning. Together with time reversal, momentum-space nonsymmorphic symmetry relates the $\Gamma$ and $M$ dispersions at opposite relative momenta, thereby protecting particle-particle nesting and the $\mathbf Q=\mathbf M$ Cooper logarithm upon doping.

We analyze the resulting interaction competition using a Wilsonian RG at charge neutrality and a full-component Bethe--Salpeter equation (BSE) at finite doping. At charge neutrality, repulsive interactions favor quantum Hall and finite-$\mathbf Q$ density-wave insulators, whereas attractive interactions yield four competing superconducting tendencies: uniform \(s^{++}\) and \(s^\pm\) states, together with spin-singlet and spin-triplet PDWs. Upon doping, momentum-space nonsymmorphic symmetry preserves the Cooper logarithm in the $\mathbf Q=\mathbf M$ PDW channel, giving it the same logarithmic enhancement as the uniform pairing channel. The finite-doping BSE reproduces the same four-region superconducting structure found in the charge-neutral RG across all sampled dopings. We further construct a \mr Chern-band model that preserves the same momentum-space nonsymmorphic symmetry and demonstrate that its leading superconducting state can be a topological PDW with a nonzero BdG Chern number.

\begin{figure}[t]
\centering
  \includegraphics[width=0.8\linewidth]{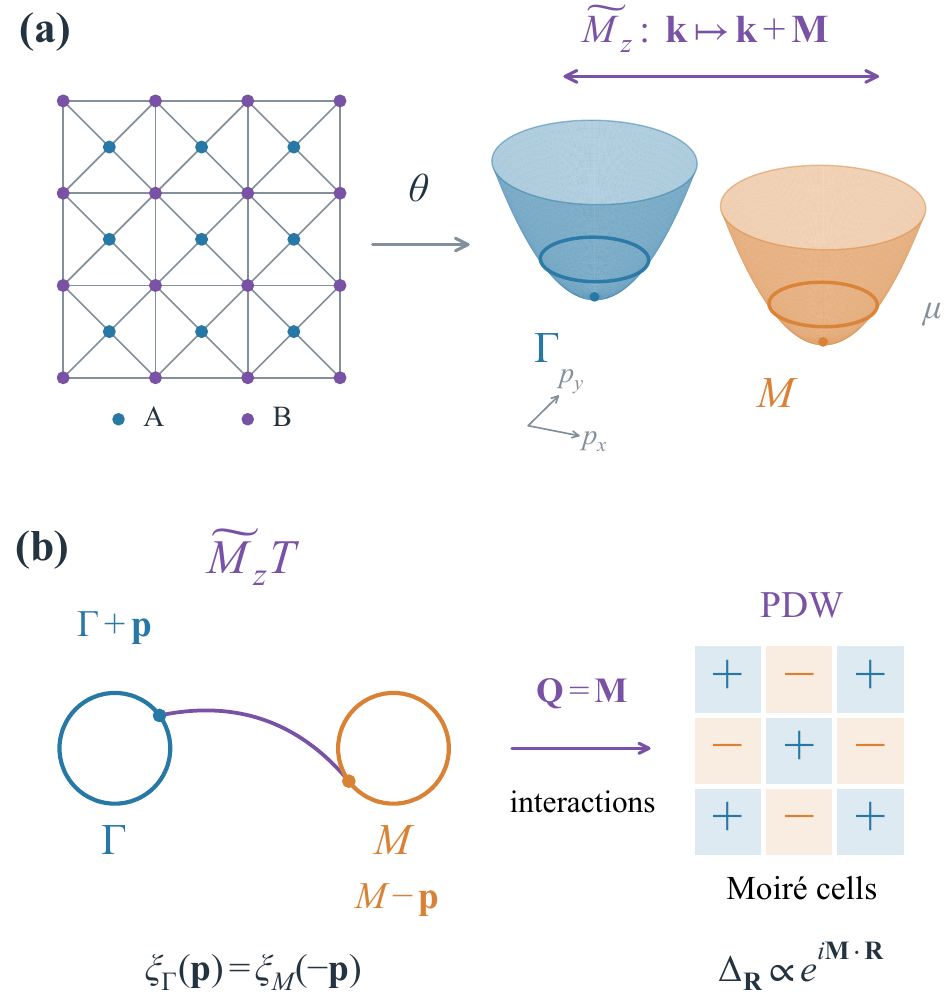}
  \caption{Nonsymmorphic particle-particle nesting and PDW pairing in the TBCB model.
(a) Single-layer checkerboard lattice and local dispersions near the symmetry-related \(\Gamma\) and \(M\) QBTs in the TBCB away from the magic-angle limit. The contours illustrate Fermi pockets generated by finite doping.
(b) The combined symmetry \(\widetilde M_z\mathcal T\) relates equal-energy states at \(\Gamma+\mathbf p\) and \(M-\mathbf p\), forming a Cooper pair with \(\mathbf Q=\mathbf M=(\pi,\pi)\). This symmetry-enforced particle-particle nesting preserves the finite-\(\mathbf Q\) Cooper logarithm. The alternating signs illustrate the resulting checkerboard modulation of the PDW phase.}
  \label{tbc}
\end{figure}

\textit{Band structure of twisted bilayer checkerboard lattice.---} The single-layer checkerboard lattice in Fig.~\ref{tbc}(a) hosts a $C_{4z}$- and time-reversal-protected QBT at $(\pi,\pi)$~\cite{PhysRevLett.103.046811}, with the general low-energy Hamiltonian:
\begin{align}
    \mathcal{H}_{0}=t_{I}(k_x^2+k_y^2)+2t_xk_xk_y\alpha_x+t_z(k_x^2-k_y^2)\alpha_z,
\end{align}
where $\alpha_{x,y,z}$ act on the sublattice space. Unlike a Dirac cone, the quadratic dispersion makes short-range four-fermion interactions tree-level marginal, permitting a systematic controlled treatment of instabilities in the weak-coupling limit.

In the TBCB geometry, twisting folds the layer QBTs to the \(\Gamma\) and \(M\) points of the mini Brillouin zone, as illustrated in Fig.~\ref{tbc}(a). Away from the magic-angle limit, these two symmetry-protected QBTs govern the low-energy weak-coupling physics. We project the microscopic \mr Hamiltonian onto the \(\Gamma\) and \(M\) QBTs, which we treat as effective valleys. In the resulting four-component low-energy spinor, \(\boldsymbol{\tau}\) acts on the \(\Gamma/M\) valley index and \(\boldsymbol{\sigma}\) on the two-component QBT spinor within each valley. In this local basis, the charge-neutral Hamiltonian is~\cite{PhysRevResearch.4.043151}
\begin{equation} 
    \mathcal{H}_{0}(\mathbf{k}) = 2t k_x k_y \gamma_2 + t(k_x^2 - k_y^2)\gamma_0, \label{eq:h0_main}
\end{equation}
where $\gamma_{0}\equiv\tau_{0}\sigma_{z}$ and $\gamma_{2}\equiv\tau_{0}\sigma_{x}$.  The valley identity $\tau_0$ in Eq.~\eqref{eq:h0_main} reflects the symmetry relation between the two QBTs. These quadratic band touchings, with their finite density of states, provide the low-energy degrees of freedom for the charge-neutral RG analysis below.

The same single-particle symmetry structure has an important consequence away from charge neutrality. In particular, the TBCB model possesses a momentum-space nonsymmorphic symmetry \(\widetilde M_z\), which relates the \(\Gamma\) and \(M\) sectors while shifting crystal momentum by \(\mathbf M\). Together with time reversal, it gives
\begin{align}
    \xi_\Gamma(\mathbf p)=\xi_M(-\mathbf p), \qquad \xi_\nu(\mathbf p) =E(\mathbf K_\nu+\mathbf p)-\mu,\label{eq:nonsym_main} 
\end{align}
with \(\mathbf K_\Gamma=0\) and \(\mathbf K_M=\mathbf M\). Upon doping, the two QBTs therefore develop symmetry-related Fermi pockets, and states at \(\Gamma+\mathbf p\) and \(M-\mathbf p\) form equal-energy Cooper pairs with total momentum \(\mathbf Q=\mathbf M\), as illustrated in Fig.~\ref{tbc}(b). This symmetry-enforced particle-particle nesting preserves the Cooper logarithm in the $\mathbf Q=\mathbf M$ channel. The explicit symmetry representations are given in the Supplemental Material.

\textit{Interactions and renormalization group analysis.---} We first consider spinless fermions at charge neutrality. In the four-component low-energy continuum theory, we must retain all symmetry-allowed four-fermion interactions. The most general local four-fermion interaction takes the form:
\begin{equation}
    \begin{aligned}
    S_{\rm int}
    &=
    \frac{1}{2} \sum_{A, B} g_{AB}
    \int d\tau d^{2} \mathbf{r}\,
    \left[\psi^{\dagger}(\mathbf{r}, \tau) \Gamma_A \psi(\mathbf{r}, \tau)\right]\\
    &\quad\times
    \left[\psi^{\dagger}(\mathbf{r}, \tau)\Gamma_B \psi(\mathbf{r}, \tau)\right].
    \end{aligned}
\label{interaction}
\end{equation}
Here $\Gamma_A$ and $\Gamma_B$ span the 16-dimensional $U(4)$ basis, consisting of the identity and the 15 generators of $SU(4)$~\cite{SupplementalMaterial}. The fermion bilinears $\psi^{\dagger}(\mathbf{r}, \tau) \Gamma_A \psi(\mathbf{r}, \tau)$ are classified by their parities under $C_{4z}$, the mirror reflections $M_{x/y}$, the nonsymmorphic rotations $\tilde C_{2x/y}$, and moir\'{e} translation $T_{\mathbf R}$, whose internal matrices are $\tau_0\sigma_y$, $\tau_0\sigma_z$, $\tau_x\sigma_z$, and $\tau_z\sigma_0$, respectively. Spinless time reversal $\mathcal T=K$ is imposed separately. In the chiral limit, particle-hole symmetry $\mathcal C=\tau_0\sigma_yK$ and chiral symmetry $\mathcal S=\tau_0\sigma_y$ constrain the single-particle spectrum without imposing additional constraints on this interaction classification.

Each matrix $\Gamma_A$ has a distinct parity set, which forbids mixed terms with $A\neq B$ in Eq.~\eqref{interaction} and leaves 16 diagonal interactions. The $SU(4)$ Fierz identities further reduce these to six independent four-fermion interactions~\cite{SupplementalMaterial}. Their RG equations, $dg_i/d\ln s=\beta_i(\{g_j\})$ with $i=1,\ldots,6$, are given in the Supplemental Material~\cite{SupplementalMaterial}.

Competing ordering tendencies are determined from the finite-scale renormalization of symmetry-breaking bilinear sources. Their one-loop flows are
\begin{equation}
    \frac{d\ln\Delta_i^X}{d\ln s}
    =2+\mathcal D_0\sum_a M^X_{ia}g_a,
    \label{eq:source_flow_main}
\end{equation}
where $X\in\{ph,pp\}$ labels the particle-hole and particle-particle channels, respectively, and $a$ runs over the six independent couplings $g_a$. The matrix $M^X$ encodes the one-loop vertex renormalization, and $\mathcal D_0$ is the QBT density of states. The particle-hole channel contains 16 independent vertices of the form $\Delta^{ph}_{A}\psi^{\dagger}\Gamma_{A}\psi$, where the $\Gamma_A$ form a basis of $4\times4$ Hermitian matrices. By contrast, Fermi statistics require the pairing matrices to be antisymmetric, $\gamma_m^T=-\gamma_m$, leaving six independent particle-particle (PP) vertices $\Delta^{pp}_{m}\psi^T\gamma_{m}\psi$. The explicit vertices and their RG flow equations are provided in the Supplemental Material.

\begin{figure}[t]
\centering
  \includegraphics[width=0.8\columnwidth]{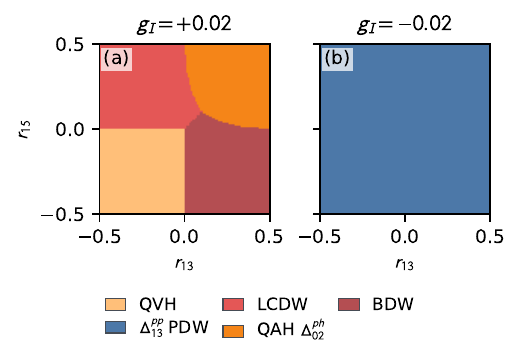}
  \caption{Spinless charge-neutral RG source maps for (a) $g_I=+0.02$ and (b) $g_I=-0.02$. The ratios $r_{13}=g_{13}/|g_I|$ and $r_{15}=g_{15}/|g_I|$ measure inter-valley exchange relative to forward scattering; band form factors make them $O(1)$ for short-range and smaller for smooth interactions. We use the representative weak-coupling window $|g_I|=0.02$ and $r_{13},r_{15}\in[-0.5,0.5]$. Colors indicate the leading RG source at the common cutoff \(\|\mathcal D_0\mathbf g(\ell_\star)\|_2=1\), determined from its integrated interaction-induced enhancement along the RG flow.}
  \label{fig:phasediagram}
\end{figure}

Projecting the double-gate screened Coulomb interaction onto the low-energy degrees of freedom provides a density-density benchmark parameterized by the intravalley forward-scattering coupling \(g_I\) and the intervalley exchange couplings \(g_{13}\) and \(g_{15}\). We explore these couplings as effective low-energy parameters, treating \(g_I<0\) as a phenomenological attraction whose microscopic origin is left unspecified.

\textit{Phase diagram of spinless fermions.---}
We summarize the RG phase diagrams of the spinless TBCB model in
Fig.~\ref{fig:phasediagram} for both repulsive and attractive intravalley
interaction $g_I$. For repulsive $g_I$, the leading instabilities include
quantum anomalous Hall (QAH) and quantum valley Hall (QVH) insulators, as well
as finite-momentum loop-current density-wave (LCDW) and bond-density-wave (BDW)
states, depending on the relative strength of intervalley couplings. The QAH
order $\Delta_{02}^{ph}=\psi^\dagger\tau_0\sigma_y\psi$ preserves moir\'{e}
translation and $C_{4z}$ but breaks time reversal and mirror symmetries. The
QVH order generates fermion masses of opposite signs at the $\Gamma$ and $M$
valleys, yielding a vanishing net charge Hall conductance. The LCDW and BDW
orders carry momentum $\mathbf Q=\mathbf M$ with order parameters
$\Delta_{3}^{ph}=\tau_x\sigma_y$ and $\Delta_{5}^{ph}=\tau_y\sigma_y$,
respectively. These orders have checkerboard patterns of loop-current and
bond-density modulation, respectively.

Remarkably, for attractive $g_I$, the PDW order $\Delta_{13}^{pp}=\tau_y\sigma_0$ with momentum $\mathbf Q=\mathbf M$ is the leading instability throughout the parameter window shown. This $C_{4z}$-invariant, locally $s$-wave pairing produces a checkerboard modulation upon condensation, while exact-band form factors may add momentum texture.

\begin{figure}[t]
\centering
  \includegraphics[width=0.8\columnwidth]{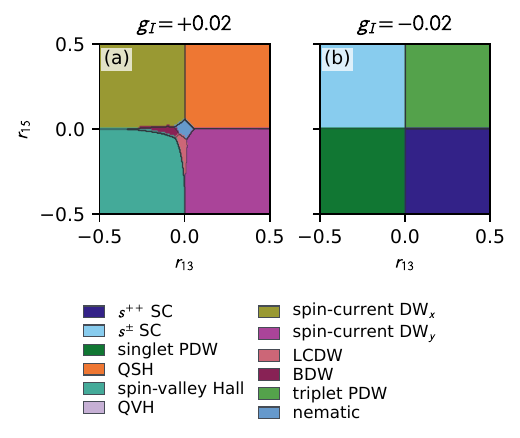}
\caption{Spin-$1/2$ charge-neutral RG source maps for (a) $g_I=+0.02$ and (b) $g_I=-0.02$, with $r_{13}=g_{13}/|g_I|$ and $r_{15}=g_{15}/|g_I|$. Colors indicate the leading RG source among all charge, spin, and singlet/triplet-pairing channels at the common cutoff $\|\mathcal D_0\mathbf g(\ell_\star)\|_2=1$; thin lines mark boundaries between leading-source regions.}\label{fig:rg}
\end{figure}

\textit{Phase diagram of spin-$1/2$ fermions.---}
We now extend the RG analysis to spin-$1/2$ fermions. In the presence of $SU(2)$ spin-rotation symmetry, each four-fermion interaction in the spinless model gives rise to two interactions, $g_{M}^c$ and $g_{M}^{s}$, in the charge and spin channels, respectively:
\begin{equation}
\begin{aligned}
&g_{M}(\psi^{\dagger}\Gamma_M\psi)(\psi^{\dagger}\Gamma_M\psi)\to\\
&g_{M}^c(\psi^{\dagger}\Gamma_Ms_{0}\psi)(\psi^{\dagger}\Gamma_Ms_{0}\psi)+g_{M}^{s}(\psi^{\dagger}\Gamma_M\vec{s}\psi)\cdot(\psi^{\dagger}\Gamma_M\vec{s}\psi),
\end{aligned}
\end{equation}
The generalized $SU(8)$ Fierz identities further reduce these 32 interactions to 16 independent interactions~\cite{SupplementalMaterial}. We take the 16 charge-charge interactions as the independent basis for the RG analysis. Meanwhile, the particle-hole (PH) channel contains 16 independent vertices in each of the charge and spin sectors. In the particle-particle (PP) channel, Fermi statistics allow 10 independent spin-singlet pairing vertices and 6 independent spin-triplet pairing vertices. The explicit forms of these vertices, together with the corresponding RG flow equations, are provided in the Supplemental Material~\cite{SupplementalMaterial}.

Assuming density-density UV interactions similar to those in the spinless case, we obtain the phase diagram in Fig.~\ref{fig:rg}. For repulsive $g_I>0$, the leading instabilities are various particle-hole insulators, including the time-reversal-symmetric quantum spin Hall (QSH) $\tau_0\sigma_y s_\alpha$ and spin-valley Hall ($\tau_z\sigma_y s_\alpha$) states, as well as momentum-$(\pi,\pi)$ spin-current density waves ($\tau_{x,y}\sigma_y s_\alpha$), LCDW ($\tau_x\sigma_y$), and BDW ($\tau_y\sigma_y$) states.

For \(g_I<0\), the complete order comparison gives four superconducting regions: uniform \(s^{++}\) \((\tau_0\sigma_0s_y)\) and \(s^\pm\) \((\tau_z\sigma_0s_y)\) superconductors, whose gaps have the same and opposite signs, respectively, on the \(\Gamma\) and \(M\) valleys, together with a spin-singlet PDW \(\tau_x\sigma_0s_y\) and a spin-triplet PDW \(\tau_y\sigma_0(i s_y s_\alpha)\). The two PDWs carry \(\mathbf Q=\mathbf M=(\pi,\pi)\), corresponding to the checkerboard modulation of the pairing order illustrated in Fig.~\ref{tbc}(b). Both are locally \(s\)-wave in the QBT orbital sector, but differ in their spin and valley-exchange structure: the singlet PDW is valley symmetric, whereas the triplet PDW is valley antisymmetric.

\textit{Bethe--Salpeter analysis at finite doping.---}
Away from charge neutrality, Eq.~\eqref{eq:nonsym_main} ensures perfect \(\Gamma\)-\(M\) particle-particle nesting at \(\mathbf Q=\mathbf M\), placing the PDW and uniform Cooper channels on the same logarithmic footing, \(\chi^{(0)}_{\rm PDW},\chi^{(0)}_{\rm SC}\sim\log(1/T)\). To determine the leading ordering tendency selected by interactions, we solve a full-component static Bethe--Salpeter equation (BSE) following an approach similar to that of \cite{PhysRevB.47.14599}. To make direct contact with the charge-neutral results discussed above, we focus on the doping regime in which the Fermi surface consists of two pockets centered at $\Gamma$ and $M$ points of the \mr Brillouin zone such that the valley and spinor indices remain well defined. We keep the effective density--density couplings $(g_I,g_{13},g_{15})$ fixed as $\mu$ is varied and antisymmetrize them to form 
the local interaction vertex.  The BSE is decomposed into physical blocks labeled by
$c=(\eta,\mathbf Q,\mathcal S)$, with $\eta=PP,PH$,
$\mathbf Q=0,\mathbf M$, and $\mathcal S$ denoting the corresponding spin
sector.  Within each block, all symmetry-allowed local order components are
retained and allowed to mix.

Let $F_A^{(c)}$ be an orthonormal basis of local order matrices in the
valley--QBT-spinor space.  The antisymmetrized
interaction vertex acts on $F_A^{(c)}$ as: 
\begin{equation*}
    \mathcal A^{PP}[F]_{\alpha\beta}
    =-\frac12 V_{\alpha\beta;\gamma\delta}F_{\gamma\delta},
    \quad
    \mathcal A^{PH}[F]_{\alpha\gamma}
    =-V_{\alpha\beta;\gamma\delta}F_{\delta\beta},
\end{equation*}
where \(\alpha,\beta,\gamma,\delta\) include spin when present. Projecting $F_A^{(c)}$ onto the exact Bloch states gives
matrix elements $X_{iA}^{(c)}$, where \(i=(\nu,\nu',\mathbf k)\)  with $\nu,\nu'$ the band indices associated with the two fermionic operators entering the order parameter. Together with the one-loop weights $\omega_i^{(c)}$, these matrix elements
 determine the PP or PH susceptibilities:
\begin{equation}
\begin{aligned}
    \Pi_{AB}^{(c)}
    &=\sum_i\omega_i^{(c)}
      X_{iA}^{(c)*}X_{iB}^{(c)},\\
    A_{AB}^{(c)}
    &=\operatorname{Tr}\!\left[
      F_A^{(c)\dagger}\mathcal A^{(c)}[F_B^{(c)}]\right].
\end{aligned}
\label{eq:main_physical_bse_matrices}
\end{equation}

\begin{figure*}[t]
  \centering
  \includegraphics[width=0.8\textwidth,keepaspectratio]
  {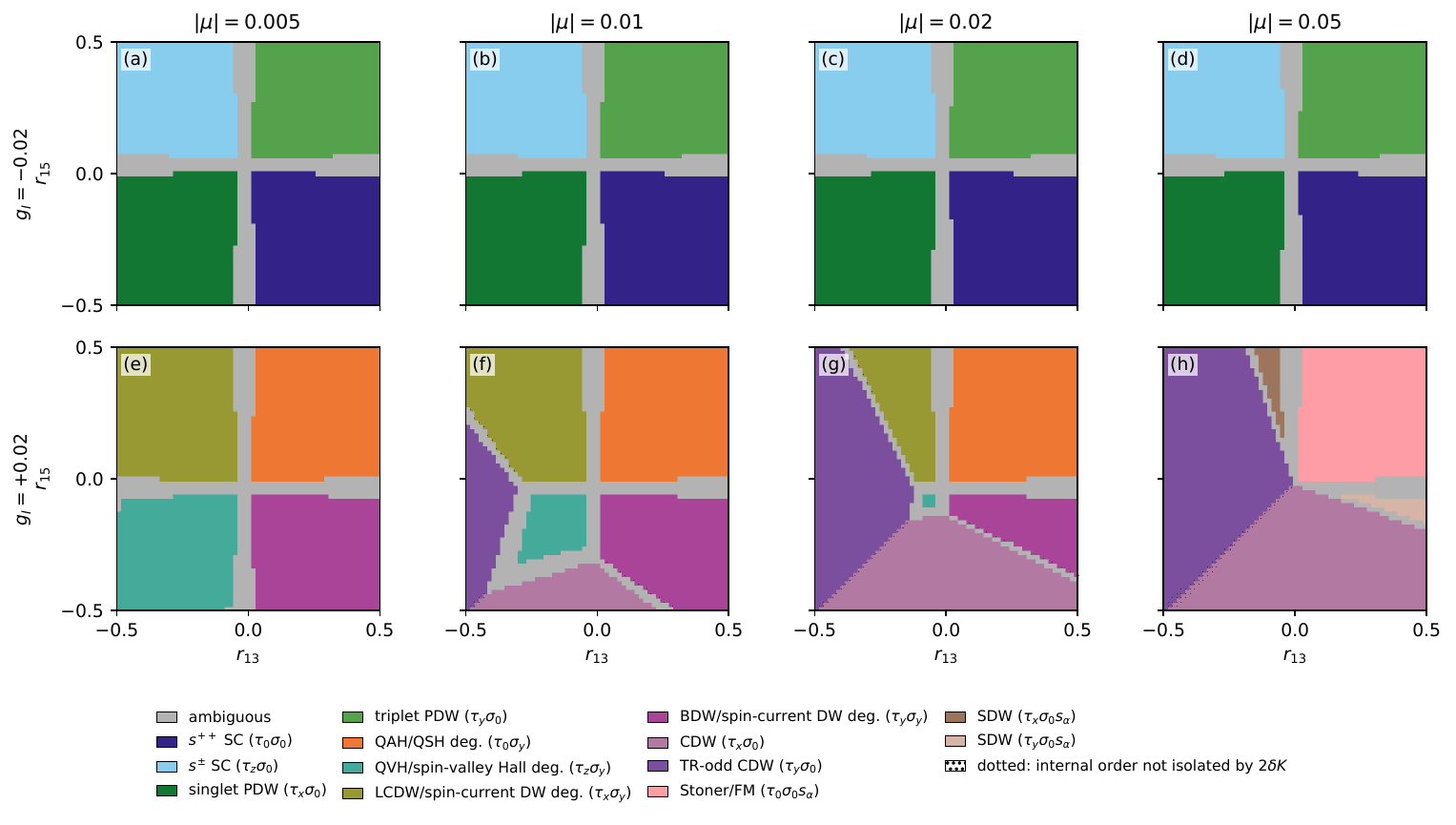}
\caption{Spinful full-component BSE at finite doping with a fixed
antisymmetrized $(g_I,g_{13},g_{15})$ vertex at
$g_I=-0.02,+0.02$ (rows) and
$|\mu|=0.005,0.010,0.020,$ and $0.050$ (columns), with
$T=0.001$ and $E_c=0.1$.  The axes are $r_{13}=g_{13}/|g_I|$ and
$r_{15}=g_{15}/|g_I|$.  Colors identify the resolved leading family and its
dominant order-matrix component among the PP and PH blocks at
$\mathbf Q=0,\mathbf M$.  Exactly degenerate charge and spin PH channels share a joint
label.  Gray points have no resolved family winner, whereas the dotted overlay
marks a resolved family whose internal order matrix is not spectrally isolated.
The quantitative assignment criteria and the corresponding spinless map are
given in the Supplemental Material.}
\label{fig:main_finite_doping_bse_spinful}
\end{figure*}

The BSE combines the available phase space encoded in $\Pi_c$ with the
interaction mixing encoded in $A_c$.  On $\operatorname{supp}\Pi_c$, we use
the Hermitian kernel
\begin{equation}
    K_c=\Pi_c^{1/2}A_c\Pi_c^{1/2},
    \qquad
    K_c y_{c,n}=\lambda_{c,n}y_{c,n}.
    \label{eq:main_physical_bse_kernel}
\end{equation}
Its largest eigenvalue $\lambda_{c,1}$ gives the leading mode within block
$c$, and comparison among blocks determines the leading BSE tendency shown in
Fig.~\ref{fig:main_finite_doping_bse_spinful}.  The corresponding local order
matrix is
$\Phi_{c,n}^{\rm loc}=\sum_A f_{c,n,A}F_A^{(c)}$, with
$f_{c,n}=\Pi_c^{-1/2}y_{c,n}$, while its momentum dependence in the exact-band
basis is carried by $X_{iA}^{(c)}$.  Further details are given in
Sec.~\ref{sec:app:finite_doping_bse} \cite{SupplementalMaterial}.

Applying this procedure, we find that for $g_I<0$ the leading BSE tendency is
predominantly in the PP sector, continuing the Cooper-channel tendency found
in the charge-neutral RG.  In the spinless case, the  PDW is favored over most of the attractive region
\cite{SupplementalMaterial}.  For spin-$1/2$ fermions, the finite-doping BSE recovers the same four superconducting families found in the charge-neutral RG: the uniform $s^{++}$ and $s^\pm$ states, together with singlet and triplet
$\mathbf Q=\mathbf M$ PDWs. At all sampled dopings, the BSE ordering maps retain the four-region superconducting structure found in the charge-neutral RG. This consistency reflects the protection of PDW provided by the momentum-space nonsymmorphic symmetry, which preserves the logarithmic Cooper enhancement of the finite-momentum pairing channels upon doping. These results therefore establish a robust weak-coupling route to both spin-singlet and spin-triplet PDW order at finite doping.

For $g_I>0$ and small $|\mu|$, the leading BSE tendency is in the PH channel and largely follows the results at
charge neutrality.  Interband transitions near the QBTs favor the
$\sigma_y$ order manifold, including the QAH/QVH channels at $\mathbf Q=0$
and the LCDW/BDW channels at $\mathbf Q=\mathbf M$.  In the spinful case,
the corresponding charge and spin partners are exactly degenerate within the
retained density--density vertex, because the direct contraction vanishes for
these $\sigma_y$ form factors.  The joint labels in
Fig.~\ref{fig:main_finite_doping_bse_spinful} therefore indicate distinct, degenerate charge and spin modes rather than mixed orders. As $|\mu|$ increases, the leading BSE tendency shifts to a charge-density-wave channel with $\mathbf Q=\mathbf M$, characterized by a form factor containing a $\sigma_0$ component. The charge–spin degeneracy is lifted in this regime.

The enhancement of PDW order by momentum-space nonsymmorphic symmetry is not specific to the TBCB model discussed above, but provides a more general mechanism that can also generate topological PDW states in \mr Chern bands. To demonstrate this, we add an \(m\alpha_y\) term to the single-particle TBCB Hamiltonian. This term breaks time-reversal symmetry but preserves both the nonsymmorphic layer-exchange symmetry \(\widetilde M_z\) and inversion \(\mathcal I\). Here inversion replaces time reversal as the momentum-reversing symmetry, so that the combined operation \(\widetilde M_z\mathcal I\) still enforces \(E_n(\mathbf k)=E_n(\mathbf M-\mathbf k)\) and hence perfect \(\mathbf Q=\mathbf M\) particle-particle nesting.  It gaps the winding-two QBTs into two Chern bands with $C=+2$ and $-2$ per spin.  At $m=0.015$, the BSE still yields the triplet $\mathbf Q=\mathbf M$ PDW as the leading instability. The quartic terms in the Ginzburg--Landau free energy further select a fully gapped unitary triplet PDW state, which realizes a topological superconducting phase with BdG Chern number $C_{\rm BdG}=8$.  This value can be understood from the (C=2) topology inherited from each gapped QBT~\cite{PhysRevResearch.4.043151}, together with the two spin copies and the particle--hole doubling intrinsic to the BdG description.

\textit{Summary and discussion.---}
We establish a weak-coupling route to PDW order in the twisted bilayer checkerboard (TBCB) model, where momentum-space nonsymmorphic symmetry together with a momentum-reversing symmetry enforces finite-\(\mathbf Q\) particle-particle nesting at generic fillings, enabling PDW order to emerge as the leading instability among competing orders. This mechanism extends to \mr Chern bands, giving rise to topological PDW states. Beyond superconductivity, the model exhibits a rich phase diagram featuring various quantum Hall phases and current- and spin-density-wave orders. More broadly, the same kinematic mechanism may operate in other systems whose momentum-space nonsymmorphic symmetries enforce finite-\(\mathbf Q\) particle-particle nesting~\cite{PhysRevLett.130.256601,Calugaru2025,htld-vgws,5zt2-scbg,bao2026moireferroelectricitydrivenbandengineering,xu2026organizingprinciplesmoirequantum}. A promising future direction is to engineer topological PDW states with an odd Bogoliubov–de Gennes (BdG) Chern number while preserving the symmetry-enforced finite-\(\mathbf Q\) sewing, thereby enabling superconducting vortices that bind an unpaired Majorana zero mode.

\textit{Note added.---}
 During the preparation of this manuscript, we became aware of an independent recent work~\cite{durrnagel2026pairdensitywaveorder} that also identifies momentum-space nonsymmorphic symmetry as a mechanism for weak-coupling PDW order through symmetry-enforced finite-momentum Cooper nesting. That work considers a projective bilayer lattice model with phonon-mediated pairing, complementary to the TBCB RG and finite-doping BSE analyses developed here.

{\it Acknowledgement:} This work is supported in part by NSFC under Grant Nos. 12347107 and 12334003 (MRL, ZW, and HY), and by the New Cornerstone Science Foundation through the Xplorer Prize (HY). Z. W. acknowledges support in part from the EPSRC under Grant No. EP/X030881/1.

\nocite{Shankar1994RG,Metzner2012FRG,Platt2013FRG,RevModPhys.63.239}
\bibliography{apssamp}

\clearpage
\onecolumngrid
\newpage

\appendix
\setcounter{secnumdepth}{3}

\startappendixlocaltoc 
\appendixlocaltoc      

\newpage

\input{supp}

\end{document}

%% file: supp.tex
\section{Review of the Twisted Bilayer Checkerboard (TBCB) Model}
In this section, we review the continuum Hamiltonian and the fundamental symmetries of the twisted bilayer checkerboard (TBCB) system, which constitute the theoretical foundation of our analysis~\cite{PhysRevResearch.4.043151}.

\subsection{Model Hamiltonian}
The TBCB system consists of two checkerboard lattices with a relative twist angle $\theta$. The isolated single-layer Hamiltonian is given by~\cite{PhysRevLett.103.046811}:
\begin{align}
    H_0\left(\mathbf{k}\right) =2t'(\cos k_{x} - \cos k_{y})\alpha_x -4t \cos \left(k_x/2\right) \cos \left(k_y/2\right)\alpha_z.
\end{align}
This lattice hosts a stable quadratic band touching (QBT) point at the Brillouin zone corner $M=(\pi, \pi)$. The corresponding low-energy $k \cdot p$ effective Hamiltonian near this QBT point reads:
\begin{align}
    H_0(\mathbf{k}) = t k_x k_y \alpha_x + t'(k_x^2 - k_y^2) \alpha_z,
\end{align}
where $\mathbf{k}=(k_x, k_y)$ is the momentum measured from the $M$ point, and $\alpha_{x, z}$ are Pauli matrices acting on the sublattice (A/B) space. We denote the two microscopic coefficients by $t_{\rm mic}\equiv t$ and $t_0\equiv t'$ and take $t_{\rm mic}=2t_0$.
The above $k\cdot p$ form uses a rotated microscopic A/B-sublattice Pauli
convention.  A direct Taylor expansion in the original microscopic convention
can differ by an overall sign and an interchange of the two anticommuting Pauli
matrices; we perform the corresponding unitary rotation once and write the
isolated-QBT Hamiltonian as
$2t_0 k_x k_y\alpha_x+t_0(k_x^2-k_y^2)\alpha_z$.  After projection to the
twisted-bilayer active bands, the $\sigma$ matrices in the low-energy theory
instead act on the symmetry-adapted QBT doublet defined below.  They reproduce
the same $k\cdot p$ matrix form but need not coincide with the microscopic
$\alpha$ matrices.  Moir\'{e} hybridization renormalizes $t_0$ to $t_{\rm eff}$;
the main text denotes this low-energy curvature by $t$.  These basis choices
do not change the spectrum or the symmetry classification, but they fix the
Dirac-matrix labels used below.

Adopting the Bistritzer-MacDonald (BM) continuum model, the low-energy effective Hamiltonian for the twisted system is $H = H_0 + H_T$. Labeling the top/bottom layers by $l=\pm 1$ (twisted by $\mp \theta/2$), we have:
\begin{align}
    H_0 &= \sum_{l=\pm 1}\sum_{\mathbf{k}} f_{l,\mathbf{k}}^{\dagger} h_{l\theta/2}(\mathbf{k}) f_{l,\mathbf{k}}
 \\ 
H_T &= \sum_{\mathbf{k}}\sum^2_{i=1}\left(f_{1,\mathbf{k}}^{\dagger}T_{i}f_{-1,\mathbf{k+q_i}} + f_{1,\mathbf{k}}^{\dagger}T_{i}f_{-1,\mathbf{k-q_i}}+\text{H.c.}\right).
\end{align}
Here, $f_{l,\mathbf{k}}$ is the fermion annihilation operator, and $h_{l\theta/2}(\mathbf{k})$ is the rotated single-layer Hamiltonian. The moir\'{e} interlayer tunneling $H_T$ is mediated by momentum transfers $\mathbf{q_1}=k_\theta \frac{(1,1)}{\sqrt{2}}$ and $\mathbf{q_2}=k_\theta \frac{(1,-1)}{\sqrt{2}}$, with the moir\'{e} wavevector $k_\theta=\frac{2\sqrt{2}\pi}{a}\sin{\frac{\theta}{2}}$. The tunneling matrices $T_i$ are parameterized by the intra- and inter-sublattice hopping amplitudes, $w_{AA}$ and $w_{AB}$:
\begin{align}
    T_1 = \begin{pmatrix} w_{AA} & w_{AB} \\ w_{AB} & w_{AA} \end{pmatrix}, \quad T_2 = \begin{pmatrix} w_{AA} & -w_{AB} \\ -w_{AB} & w_{AA} \end{pmatrix}.
\end{align}
In the chiral limit, defined by vanishing intra-sublattice tunneling ($w_{AA}=0$), exactly flat bands with nontrivial topology emerge at a discrete series of magic angles~\cite{PhysRevResearch.4.043151}.

\begin{figure}
    \centering
    \includegraphics[width=0.7\linewidth]{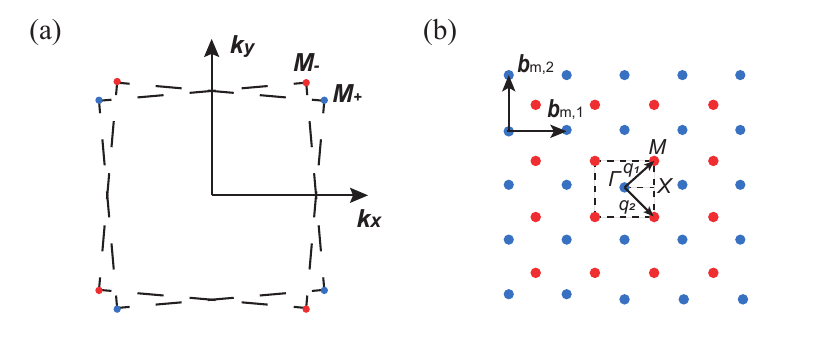}
\caption{MiniBZ of the TBCB model
}
    \label{fig:minibz}
\end{figure}

The interlayer tunneling defines the reciprocal vectors of the moir\'{e} Brillouin zone (mBZ):
\begin{align}
    \mathbf{b}_{m,1}=\sqrt{2}k_\theta(1,0),\quad \mathbf{b}_{m,2}=\sqrt{2}k_\theta(0,1),
\end{align}
which span the momentum-space lattice $\mathcal{Q}_l=\mathbf{Q}_{0,l}+\mathbb{Z}\mathbf{b}_{m,1}+\mathbb{Z}\mathbf{b}_{m,2}$ with $\mathbf{Q}_{0,+}=\mathbf{0}$ and $\mathbf{Q}_{0,-}=\mathbf{q}_1$, as illustrated in Fig.~\ref{fig:minibz}. In the ensuing analysis, we adopt the $\mathcal{Q}$-lattice plane-wave basis $f_{l,\mathbf{Q}_l}(\mathbf{k})=f_{l,\mathbf{k-Q_l}}$.

\subsection{Symmetries of the TBCB Model}\label{app:sec:tbcb_sym}
The low-energy physics of the TBCB Hamiltonian is strictly constrained by a set of discrete symmetries. Let $\alpha_{i=x,y,z}$ denote the Pauli matrices acting on the microscopic sublattice degree of freedom, while $l=\pm1$ labels the layer.

We first summarize the standard crystalline and non-spatial symmetries identified in Ref.~\cite{PhysRevResearch.4.043151}:
\begin{itemize}
    \item \textbf{Mirror Symmetry ($M_{x/y}$)}: The system is invariant under mirror reflections about the $x$ or $y$ plane, acting on the field operators as: 
    \begin{align}
            M_{x/y} f_{l,\mathbf{Q}_l}(\mathbf{k}) M_{x/y}^{-1}= \alpha_{z} f_{l,M_{x/y} \mathbf{Q}_l}(M_{x/y}\mathbf{k}).
    \end{align}
     The Hamiltonian trivially commutes with this operation: $\left[H, M_{x/y}\right]=0$.
    \item \textbf{$C_{4z}$ Rotational Symmetry}: The model preserves a four-fold rotation symmetry about the $z$-axis:
    \begin{align}
        C_{4z} f_{l,\boldsymbol{Q}_l}(\mathbf{k}) C_{4z}^{-1}=\alpha_{y} f_{l,R_{\pi/2} \boldsymbol{Q}_l}(R_{\pi/2}\mathbf{k}),
    \end{align}
    where $R_{\pi/2}$ denotes a $\pi/2$ spatial rotation acting on $\mathbf{k}$, satisfying $\left[H, C_{4z}\right]=0$.
    \item \textbf{Time-Reversal Symmetry (TRS)}: Assuming a spinless basis, TRS is represented by complex conjugation $\mathcal T=K$. Its action is:
    \begin{align}
    \mathcal T f_{l,\boldsymbol{Q}_l}(\mathbf{k}) \mathcal T^{-1}=f_{l,-\boldsymbol{Q}_l}(-\mathbf{k}).
    \end{align}
    The commutation $\left[H,\mathcal T\right]=0$ dictates that the single-particle Hamiltonian satisfies $h_{\mathbf{Q}_l,\mathbf{Q}^{\prime}_l}(\mathbf{k})=h_{-\mathbf{Q}_l,-\mathbf{Q}^{\prime}_l}^*(-\mathbf{k})$.
    \item \textbf{Particle-Hole Symmetry (PHS)}: In the chiral limit ($w_{AA} = 0$), the model acquires an exact particle-hole transformation $P$:
    \begin{align}
        Pf_{l,\boldsymbol{Q}_l}(\mathbf{k})P^{-1}=\alpha_{y} f_{l,-\boldsymbol{Q}_l}(-\mathbf{k})^{\dagger}.
    \end{align}
    In first quantization, the corresponding PHS operator is antiunitary,
    $\mathcal C=U_CK$, with $U_C=\alpha_y$, and obeys
    \begin{align}
        U_C h^*(-\mathbf{k})U_C^\dagger=-h(\mathbf{k}).
    \end{align}
    A finite $w_{AA}$ explicitly breaks this symmetry.
    \item \textbf{Chiral Symmetry}: Combining spinless TRS, $\mathcal T=K$, with PHS gives the unitary chiral operator
    \begin{align}
        \mathcal S=\mathcal C\mathcal T=U_C=\alpha_y,
        \qquad
        \mathcal S h(\mathbf{k})\mathcal S^{-1}=-h(\mathbf{k}).
    \end{align}
    Thus $\{\mathcal S,h(\mathbf{k})\}=0$ in the chiral limit.
\end{itemize}

The TBCB model also preserves emergent symmetries that act nonsymmorphically in momentum space. Unlike standard point-group symmetries, which map $\mathbf{k}$ to $R\mathbf{k}$, these operations include an irreducible momentum shift that cannot be gauged away.

Specifically, we identify effective in-plane two-fold rotations, $\tilde{C}_{2x}$ and $\tilde{C}_{2y}$, which couple the layer exchange to a translation in the moir\'{e} momentum lattice:
\begin{align}
\tilde{C}_{2x} f_{l, \mathbf{Q}_l}(\mathbf{k}) \tilde{C}_{2x}^{-1} = \alpha_z  f_{-l, C_{2x} \mathbf{Q}_l - \mathbf{q}_{2}}(C_{2x} \mathbf{k} - \mathbf{q}_{2}), \quad \tilde{C}_{2y} f_{l, \mathbf{Q}_l}(\mathbf{k}) \tilde{C}_{2y}^{-1} =\alpha_z  f_{-l, C_{2y} \mathbf{Q}_l + \mathbf{q}_{2}}(C_{2y} \mathbf{k} + \mathbf{q}_{2}).
\end{align}
In the microscopic layer--sublattice tensor-product space, this operation exchanges the two layers and acts with $\alpha_z$ on the sublattice degree of freedom. Because the two layer Brillouin-zone centers differ, $l\to-l$ also shifts the moir\'{e} momentum. This is the reciprocal-space analogue of a real-space nonsymmorphic symmetry.

Combining either momentum-shifting rotation with the corresponding mirror defines $\tilde M_z=M_x\tilde C_{2y}=M_y\tilde C_{2x}$, with
\begin{align}
\tilde M_z f_{l,\mathbf Q_l}(\mathbf k)\tilde M_z^{-1}
=f_{-l,\mathbf Q_l-\mathbf q_1}(\mathbf k-\mathbf q_1).
\end{align}
Here $\mathbf q_1=(\mathbf b_{m,1}+\mathbf b_{m,2})/2\equiv\mathbf M$ modulo a moir\'e reciprocal vector, so $E_{\nu}(\mathbf k)=E_{\nu}(\mathbf k-\mathbf M)$. Combining this relation with time reversal gives
\begin{equation}
E_{\nu}(\Gamma+\mathbf p)=E_{\nu}(M-\mathbf p),
\qquad
\xi_{\nu,\Gamma}(\mathbf p)=\xi_{\nu,M}(-\mathbf p).
\label{eq:app_ns_pdw_nesting}
\end{equation}
Equation~\eqref{eq:app_ns_pdw_nesting} enforces particle-particle nesting whenever these bands cross the Fermi level; the interaction vertex and Bloch form factors select the spin and orbital component.

\section{Symmetry-Allowed Interactions}

In this section, we systematically classify the interaction channels compatible with the system's symmetry group. To explicitly construct the low-energy effective field theory, we project the microscopic moir\'{e} Hamiltonian onto the relevant low-energy subspace. Away from the magic angle, the band dispersion reveals that the low-energy physics is governed by two QBT points located at the moir\'{e} Brillouin zone center $\Gamma$ and corner $M$, which we identify as two effective ``valleys''. Throughout this section, ``general'' and ``complete'' refer to momentum-independent, zero-derivative interactions in the retained four-component low-energy basis.

The retained spinor carries a valley index $\tau=\Gamma,M$, which labels patches around the two QBTs, and a two-component QBT index $\sigma$ within each patch. We first consider spinless fermions and give the spin-$\frac12$ generalization below. The numerical basis spanning the two active bands has a momentum-dependent \(U(2)\) freedom, which can make the corresponding symmetry matrices momentum dependent. A local frame independent of this momentum-dependent numerical gauge follows from the rank-two spectral projector \(P_\tau(\mathbf p)\) onto these bands:
\begin{equation}
    V_{\tau=\Gamma,\mathbf{M}}(\mathbf{p})= P_{\tau}(\mathbf{p}) W_{\tau} [W^{\dagger}_{\tau}P_{\tau}(\mathbf{p})W_{\tau}]^{-\frac{1}{2}},
\end{equation}
where the orthonormal columns of $W_\tau$ span the QBT subspace at $\mathbf p=0$. The column labels of $V_\tau$ and $W_\tau$ define $\sigma$. With $A_\tau(\mathbf p)=W_\tau^\dagger P_\tau(\mathbf p)W_\tau$, the frame is well defined wherever $A_\tau(\mathbf p)$ is positive definite. Since $P_\tau^2=P_\tau$,
\begin{equation}
    V_\tau^\dagger(\mathbf p)V_\tau(\mathbf p)
    =A_\tau^{-1/2}(\mathbf p)A_\tau(\mathbf p)A_\tau^{-1/2}(\mathbf p)
    =\sigma_0,
\end{equation}
and $V_\tau(0)=W_\tau$. The projector is unchanged by momentum-dependent $U(2)$ rotations of the numerical eigenvectors, and hence so is $V_\tau$.

For $g=M_{x/y},C_{4z},\tilde C_{2x/y},\tilde M_z$, a momentum $\mathbf K_\tau+\mathbf p$ in patch $\tau$ is mapped to $\mathbf K_{g\tau}+R_g\mathbf p$ modulo a moir\'e reciprocal vector. With the fixed reciprocal-lattice, each operation is represented in the microscopic plane-wave basis by a momentum-independent unitary map $D_{g,\tau}$, which includes the layer and sublattice action together with the corresponding relabeling of plane-wave indices. The active-band projectors therefore satisfy
\begin{equation}
P_{g\tau}(R_g\mathbf p)
=D_{g,\tau}P_\tau(\mathbf p)
D^{\dagger}_{g,\tau}.
\label{eq:projector_symmetry_covariance}
\end{equation}
Since $P_\tau(\mathbf p)V_\tau(\mathbf p)=V_\tau(\mathbf p)$,
Eq.~\eqref{eq:projector_symmetry_covariance} gives
\begin{equation}
P_{g\tau}(R_g\mathbf p)
D_{g,\tau}V_\tau(\mathbf p)
=D_{g,\tau}V_\tau(\mathbf p).
\end{equation}
Thus $D_{g,\tau}V_\tau(\mathbf p)$ lies in the target active subspace and can be expanded in the orthonormal frame $V_{g\tau}(R_g\mathbf p)$ as
\begin{align}
D_{g,\tau}V_\tau(\mathbf p)
&=V_{g\tau}(R_g\mathbf p)
G_{g;g\tau,\tau}(\mathbf p),
\label{eq:sewing_frame_expansion}\\
G_{g;g\tau,\tau}(\mathbf p)
&=V_{g\tau}^{\dagger}(R_g\mathbf p)
D_{g,\tau}V_\tau(\mathbf p).
\label{eq:exact_low_energy_sewing}
\end{align}
Using projector covariance and the orthonormality of the local frames,
\begin{align}
G_{g;g\tau,\tau}^{\dagger}(\mathbf p)
G_{g;g\tau,\tau}(\mathbf p)
&=
V_\tau^\dagger(\mathbf p) D^{\dagger}_{g,\tau}
P_{g\tau}(R_g\mathbf p) D_{g,\tau} V_\tau(\mathbf p) \nonumber\\
&=
V_\tau^\dagger(\mathbf p)P_\tau(\mathbf p)
V_\tau(\mathbf p)=\sigma_0,
\end{align}
so $G_{g;g\tau,\tau}(\mathbf p)$ is a unitary $2\times2$ sewing matrix.

We now fix the remaining constant gauge freedom in the reference bases \(W_\tau\) so that the symmetry representations take simple forms. Since $V_\tau(0)=W_\tau$, evaluating Eqs.~\eqref{eq:sewing_frame_expansion} and \eqref{eq:exact_low_energy_sewing} at $\mathbf p=0$ gives
\begin{equation}
\rho_{g,\tau}
\equiv G_{g;g\tau,\tau}(0)=W_{g\tau}^{\dagger}D_{g,\tau}W_\tau,
\qquad
D_{g,\tau}W_\tau=W_{g\tau}\rho_{g,\tau}.
\label{eq:anchor_symmetry_representation}
\end{equation}
We construct $W_\Gamma$ by diagonalizing $M_x$ within the $\Gamma$-point QBT subspace and ordering its eigenvectors by mirror eigenvalues $(+1,-1)$, so that
$\rho_{M_x,\Gamma}=\sigma_z$. Using the remaining relative phase freedom, we choose
$\rho_{C_{4z},\Gamma}=\sigma_y$. With the microscopic symmetry-operator phases chosen in Appendix~\ref{app:sec:tbcb_sym}, this also gives $\rho_{M_y,\Gamma}=\sigma_z$. The choice
$W_M=D_{\tilde M_z,\Gamma}W_\Gamma$ carries the same component ordering to $M$ and fixes the relative valley gauge, so that \(\widetilde M_z\) has representation \(\tau_x\) in valley space.

Projector covariance together with Eq.~\eqref{eq:anchor_symmetry_representation} gives
\begin{align}
A_{g\tau}(R_g\mathbf p)
&=\rho_{g,\tau}A_\tau(\mathbf p)\rho_{g,\tau}^\dagger,\nonumber\\
A_{g\tau}(R_g\mathbf p)^{-1/2}
&=\rho_{g,\tau}A_\tau(\mathbf p)^{-1/2}\rho_{g,\tau}^\dagger,
\label{eq:anchor_overlap_symmetry_covariance}
\end{align}
where the second identity follows from the unique positive square root. Substituting these relations into the definition of $V_\tau$ yields
\begin{align}
V_{g\tau}(R_g\mathbf p)
&=D_{g,\tau}V_\tau(\mathbf p)\rho_{g,\tau}^\dagger,\nonumber\\
G_{g;g\tau,\tau}(\mathbf p)
&=\rho_{g,\tau}.
\label{eq:constant_local_sewing}
\end{align}
Thus the sewing matrices are momentum independent throughout the QBT patches wherever $A_\tau(\mathbf p)$ is positive definite.

Having fixed the momentum-independent sewing blocks $\rho_{g,\tau}$, we assemble them into the full internal representation on the ordered spinor
$(\Gamma,1;\Gamma,2;M,1;M,2)$:
\begin{equation}
G_g^{\rm int}
=\sum_{\tau=\Gamma,M}
|g\tau\rangle\langle\tau|\otimes\rho_{g,\tau}.
\label{eq:full_internal_symmetry_from_sewing}
\end{equation}
With the relative valley gauge fixed above, the mirrors and $C_{4z}$ preserve each valley and have blocks $\sigma_z$ and $\sigma_y$, respectively, at both QBTs. The choice
$W_M=D_{\tilde M_z,\Gamma}W_\Gamma$ gives $\tilde M_z$ the identity block between the two valleys. The relations
$\tilde M_z=M_x\tilde C_{2y}=M_y\tilde C_{2x}$ then fix the QBT block of the valley-exchanging $\tilde C_{2x/y}$ to $\sigma_z$. A primitive moir\'e translation acts as $\sigma_0$ on the QBT index and contributes phases $+1$ and $-1$ at $\Gamma$ and $M$, respectively. Writing $G_g\equiv G_g^{\rm int}$, Eq.~\eqref{eq:full_internal_symmetry_from_sewing} therefore gives
\begin{equation}
G_{\tilde C_{2x/y}}=\tau_x\sigma_z,\qquad
G_{M_{x/y}}=\tau_0\sigma_z,\qquad
G_{C_{4z}}=\tau_0\sigma_y,\qquad
G_{T_{\mathbf R}}=\tau_z\sigma_0.
\label{eq:symmetry_parity_generators}
\end{equation}
Here $T_{\mathbf R}$ denotes either primitive translation,
$\mathbf R=\mathbf a_1$ or $\mathbf a_2$. For a general lattice vector
$\mathbf R=m\mathbf a_1+n\mathbf a_2$, its internal matrix is
$\operatorname{diag}[1,(-1)^{m+n}]\otimes\sigma_0$. With the Bloch-phase
convention $e^{i\mathbf k\cdot\mathbf R}$, the full translation sewing matrix
contains a common factor $e^{i\mathbf p\cdot\mathbf R}$ describing the spatial
translation of the envelope field, together with the internal matrix above.
Overall symmetry phases cancel in the conjugation of charge-neutral local
bilinears.

There are 16 basis matrices for $4\times4$ Hermitian matrices, listed in
Table~\ref{tab:symmetry_classification}. For each basis matrix $\Gamma_A$, we define
$$
G_g\Gamma_A G_g^{-1}=\eta_g\Gamma_A,
$$
with $\eta_g=+1$ ($-1$) for commutation (anticommutation). The 16 Hermitian
basis matrices occupy all 16 distinct parity sectors under the four operations,
so each bilinear is assigned a unique parity tuple, as summarized in
Table~\ref{tab:symmetry_classification}.

\begin{table}[h]
\centering
\renewcommand{\arraystretch}{1.2} 
\begin{tabular}{cc|cccc}
\hline \hline
\multirow{2}{*}{Matrix} & \multirow{2}{*}{Representation ($\tau \otimes \sigma$)} & \multicolumn{4}{c}{Symmetry Parity $\eta$} \\
 &  & $\tilde{C}_{2x/y}$ & $M_{x/y}$ & $C_{4z}$ & $T_{\mathbf{R}}$ \\ \hline
$\mathbb{I}$    & $\tau_0 \sigma_0$ & $+$ & $+$ & $+$ & $+$ \\ 
$\gamma_{15}$   & $\tau_x \sigma_0$ & $+$ & $+$ & $+$ & $-$ \\ 
$\gamma_{0}$    & $\tau_0 \sigma_z$ & $+$ & $+$ & $-$ & $+$ \\ 
$\gamma_{23}$   & $\tau_x \sigma_z$ & $+$ & $+$ & $-$ & $-$ \\ 
$\gamma_{1}$    & $\tau_z \sigma_y$ & $+$ & $-$ & $+$ & $+$ \\ 
$\gamma_{5}$    & $\tau_y \sigma_y$ & $+$ & $-$ & $+$ & $-$ \\ 
$\gamma_{01}$   & $\tau_z \sigma_x$ & $+$ & $-$ & $-$ & $+$ \\ 
$\gamma_{05}$   & $\tau_y \sigma_x$ & $+$ & $-$ & $-$ & $-$ \\ 
$\gamma_{35}$   & $\tau_z \sigma_0$ & $-$ & $+$ & $+$ & $+$ \\ 
$\gamma_{13}$   & $\tau_y \sigma_0$ & $-$ & $+$ & $+$ & $-$ \\ 
$\gamma_{12}$   & $\tau_z \sigma_z$ & $-$ & $+$ & $-$ & $+$ \\ 
$\gamma_{25}$   & $\tau_y \sigma_z$ & $-$ & $+$ & $-$ & $-$ \\ 
$\gamma_{02}$   & $\tau_0 \sigma_y$ & $-$ & $-$ & $+$ & $+$ \\ 
$\gamma_{3}$    & $\tau_x \sigma_y$ & $-$ & $-$ & $+$ & $-$ \\ 
$\gamma_{2}$    & $\tau_0 \sigma_x$ & $-$ & $-$ & $-$ & $+$ \\ 
$\gamma_{03}$   & $\tau_x \sigma_x$ & $-$ & $-$ & $-$ & $-$ \\ 
\hline \hline
\end{tabular}
\caption{Symmetry classification of the 16 Hermitian basis matrices. The parity eigenvalues $\eta = \pm 1$ (denoted as $+$ and $-$) are determined by the commutation ($+$) or anticommutation ($-$) relations with the internal generators $G_{\tilde C_{2x/y}}$, $G_{M_{x/y}}$, $G_{C_{4z}}$, and $G_{T_{\mathbf R}}$ in Eq.~\eqref{eq:symmetry_parity_generators}. The Pauli matrix representation $\tau \otimes \sigma$ is included for reference.}
\label{tab:symmetry_classification}
\end{table}

Because each Hermitian matrix $\Gamma_A$ belongs to a unique parity sector defined by $(\eta_{\tilde{C}_{2x/y}}, \eta_{M_{x/y}}, \eta_{C_{4z}}, \eta_{T_{\mathbf{R}}})$, a mixed interaction $(\psi^{\dagger}\Gamma_A\psi)(\psi^{\dagger}\Gamma_B\psi)$ with $A \neq B$ is not invariant under all four spatial operations.  The spatially invariant contact action therefore contains the 16 diagonal terms $(\psi^{\dagger}\Gamma_A\psi)^2$.  The chiral-limit PHS $\mathcal C=\sigma_yK$ and chiral operator $\mathcal S=\sigma_y$ are single-particle spectral constraints rather than additional spatial generators in Table~\ref{tab:symmetry_classification}.  Although $\mathcal S$ and $C_{4z}$ have the same internal matrix $\sigma_y$, their momentum-space actions differ: $\mathcal S$ anticommutes with $h(\mathbf k)$ at fixed momentum, whereas $C_{4z}$ commutes with the Hamiltonian after rotating $\mathbf k$.  To check the quartic terms, we understand the bilinears as normal ordered relative to charge neutrality.  For $B_A=\psi^\dagger\Gamma_A\psi$,
\begin{equation}
    \mathcal S B_A\mathcal S^{-1}
    =\psi^\dagger\sigma_y\Gamma_A\sigma_y\psi,
    \qquad
    P\,{:}B_A{:}\,P^{-1}
    =-{:}\psi^\dagger\sigma_y\Gamma_A\sigma_y\psi{:}.
    \label{eq:phs_chiral_bilinear_action}
\end{equation}
Since $\sigma_y\Gamma_A\sigma_y=\pm\Gamma_A$ for every Pauli-product basis matrix, both operations leave $({:}B_A{:})^2$ invariant.  For the identity matrix, normal ordering removes the constant and one-body pieces generated by $n\mapsto4-n$.  Thus PHS and chiral symmetry impose no further relation among the 16 diagonal quartic couplings before the Fierz reduction.  The most general symmetry-allowed interaction action is therefore
\begin{equation}
    \begin{aligned}
    S_{\rm int}=&\frac{1}{2} \int d\tau d^{2} \mathbf{r}\Big[g_{0}\left(\psi^{\dagger}\gamma_{0}\psi\right)^2+g_{01}\left(\psi^{\dagger}\gamma_{01}\psi\right)^2+g_{2}\left(\psi^{\dagger}\gamma_{2}\psi\right)^2+g_{12}\left(\psi^{\dagger}\gamma_{12}\psi\right)^2\\
    +&g_{1}\left(\psi^{\dagger}\gamma_{1}\psi\right)^2+g_{02}\left(\psi^{\dagger}\gamma_{02}\psi\right)^2+g_{I}\left(\psi^{\dagger}\psi\right)^2+g_{35}\left(\psi^{\dagger}\gamma_{35}\psi\right)^2\\
    +&g_{3}\left(\psi^{\dagger}\gamma_{3}\psi\right)^2+g_{5}\left(\psi^{\dagger}\gamma_{5}\psi\right)^2+g_{03}\left(\psi^{\dagger}\gamma_{03}\psi\right)^2+g_{05}\left(\psi^{\dagger}\gamma_{05}\psi\right)^2\\
    +&g_{13}\left(\psi^{\dagger}\gamma_{13}\psi\right)^2+g_{15}\left(\psi^{\dagger}\gamma_{15}\psi\right)^2+g_{23}\left(\psi^{\dagger}\gamma_{23}\psi\right)^2+g_{25}\left(\psi^{\dagger}\gamma_{25}\psi\right)^2\Big].
    \end{aligned}\label{equ:15interactions}
\end{equation}
For future convenience, we adopt $V_{i}$ to denote the four-fermion interaction term corresponding to the coupling parameter $g_i$ in Eq.~\eqref{equ:15interactions}.

The dimensionality of this interaction space can be further strictly constrained by exploiting the Fierz identities inherent to the four-component spinor formalism. The general $SU(4)$ Fierz identity for a spinless system is given by:
\begin{equation}
\left(\psi^{\dagger} M \psi\right)\left(\psi^{\dagger} N \psi\right) =-\frac{1}{16}\operatorname{Tr}\left[ M \Gamma_{A} N \Gamma_{B}\right]\left(\psi^{\dagger} \Gamma_{B} \psi\right)\left(\psi^{\dagger} \Gamma_{A} \psi\right),\label{eq:fierz_spinless}
\end{equation}
where $M$ and $N$ are arbitrary $4\times4$ Hermitian matrices, and the summation indices $A$ and $B$ run over the complete set of 16 basis matrices defined previously.

We apply this identity to each of the symmetry-allowed interaction terms, denoted as $\mathcal{O}_i = (\psi^\dagger \Gamma_i \psi)^2$. The expansion yields:
\begin{align} 
\mathcal{O}_i &= -\frac{1}{16} \sum_{j,k} \text{Tr}[\Gamma_i \Gamma_k \Gamma_i \Gamma_j] (\psi^\dagger \Gamma_j \psi)(\psi^\dagger \Gamma_k \psi) \nonumber \\ 
&= -\frac{1}{16} \sum_j \text{Tr}[\Gamma_i \Gamma_j \Gamma_i \Gamma_j] \mathcal{O}_{j}. 
\end{align}
The second line retains only $j=k$ because the distinct parity tuples in Table~\ref{tab:symmetry_classification} exclude mixed products $(\psi^\dagger \Gamma_j \psi)(\psi^\dagger \Gamma_k \psi)$ with $j\neq k$.

This system of constraints can be recast as a linear matrix equation $\mathcal{F}\mathbf{V}=0$, where $\mathbf{V}$ is the 16-dimensional vector spanned by the allowed quartic interaction terms $\mathcal{O}_i$. The matrix $\mathcal{F}$ encapsulates the linear dependencies imposed by the Clifford algebra. An analysis of the null space of $\mathcal{F}$ reveals that the number of linearly independent interaction terms fundamentally reduces from sixteen to six. Before proceeding to the RG flow, we project a microscopic scalar interaction onto this 4-dimensional low-energy spinor basis to define a bare density-density benchmark and the associated initial couplings.

\subsection{Projected Coulomb Interaction}\label{sec:app:projected_interaction}

We obtain a bare density-density benchmark for the low-energy model by projecting the microscopic Coulomb interaction onto the active bands near the
$\Gamma$ and $M$ QBTs.  For a double-gate geometry, the screened Coulomb interaction is~\cite{PhysRevB.103.205414}
\begin{align}
V(\qq) = \frac{2\pi e^2}{\epsilon q} \tanh\left(\frac{q d}{2}\right),
\end{align}
where $d$ is the gate distance and $\epsilon$ is the dielectric constant.  We
use $d=10$ nm and $\epsilon=10$.

The normal-ordered interaction Hamiltonian is
\begin{align}
H_{I} = \frac{1}{2\Omega_{tot}} \sum_{\GG,\qq} V(\qq+\GG) :\rho_{-\qq-\GG} \rho_{\qq +\GG}:
\end{align}
where $\Omega_{tot}$ is the total system area and $\GG$ is a moir\'{e}
reciprocal lattice vector.  Let $r,s\in\{1,2\}$ label the two active bands.
Projection of the density operator gives
\begin{align}
\rho_{\qq+\GG} \approx \sum_{\kk, r, s} \Lambda_{rs}(\kk, \qq+\GG) \cre{\psi}{\kk, r} \des{\psi}{\kk+\qq, s}.
\end{align}
Here $\kk$ lies in the first moir\'{e} Brillouin zone (mBZ), and the Bloch
form factor is
\begin{align}
\Lambda_{rs}(\kk, \qq+\GG) = \langle \psi_{\kk,r} | e^{-i(\qq+\GG)\cdot \hat{\mathbf{r}}} | \psi_{\kk+\qq,s} \rangle = \sum_{\mathbf{Q}} u^{ *}_{r, \mathbf{Q}}(\kk) u_{s, \mathbf{Q}-\GG}(\kk+\qq),
\end{align}
where $u_{s,\mathbf Q}(\kk)$ is the plane-wave coefficient of
$|\psi_{r,\kk}\rangle$.

Substitution into $H_I$ gives the band-projected interaction
\begin{align}
H_I \approx \frac{1}{2\Omega_{tot}} \sum_{\qq, \GG} \sum_{\kk, \kk'} \sum_{r,s,t,u} V(\qq+\GG)\left[ \Lambda_{rs}(\kk, \qq+\GG) \cre{\psi}{\kk, r} \des{\psi}{\kk+\qq, s} \right]\left[ \Lambda_{tu}(\kk', -\qq-\GG) \cre{\psi}{\kk', t} \des{\psi}{\kk'-\qq, u} \right].
\end{align}
Using
$\Lambda_{tu}(\kk',-\qq-\GG)=\Lambda_{ut}^*(\kk'-\qq,\qq+\GG)$,
this becomes the standard four-fermion form
\begin{align}
H_I = \frac{1}{2N} \sum_{\qq, \kk, \kk'} \sum_{r,s,t,u} \mathcal{V}_{rstu}(\qq, \kk, \kk')\cre{\psi}{\kk, r} \cre{\psi}{\kk', t} \des{\psi}{\kk'-\qq, u} \des{\psi}{\kk+\qq, s}\label{app:eq:interaction_momentum},
\end{align}
with the effective interaction tensor $\mathcal{V}_{rstu}$
\begin{align}
\mathcal{V}_{rstu}(\qq, \kk, \kk') = \frac{1}{\Omega_0} \sum_{\GG} V(\qq+\GG) \Lambda_{rs}(\kk, \qq+\GG) \Lambda_{ut}^*(\kk'-\qq, \qq+\GG).
\end{align}

To derive the parameters for the effective field theory, we focus on the low-energy physics strictly bounded around the $\Gamma$ and $M$ points. We label the resulting four-component local states by
$a=(v,\alpha)$, where $v=\Gamma,M$ is the valley and $\alpha=1,2$ is the QBT
spinor index.  Evaluating $\mathcal V_{rstu}$ at the corresponding anchor
momenta defines the local density-bilinear tensor.  For
$a=(v_a,\alpha_a)$, $b=(v_b,\alpha_b)$, $c=(v_c,\alpha_c)$, and
$d=(v_d,\alpha_d)$, the projection is
\begin{equation*}
\mathcal V^{(0)}_{ab;cd}
=\mathcal V_{\alpha_a\alpha_b\alpha_c\alpha_d}
\!\left(\mathbf K_{v_b}-\mathbf K_{v_a},
\mathbf K_{v_a},\mathbf K_{v_c}\right),
\end{equation*}
for valley combinations satisfying momentum conservation modulo a reciprocal
lattice vector.  Thus an intravalley bilinear has zero transfer, whereas an
intervalley bilinear carries $\mathbf K_M-\mathbf K_\Gamma$ or its reverse.

We identify the $\Gamma$ and $M$ QBTs with $\tau_z=+1$ and $\tau_z=-1$,
respectively, and expand the local interaction in the 16 Hermitian matrices
$T^i$ acting in valley--QBT-spinor space:
\begin{align} 
H_{\text{eff}} = \frac{1}{2N} \sum_{\kk, \kk', \qq} \sum_{i} g_i \left[ \sum_{a,b} (T^i)_{ab} \cre{\psi}{\kk, a} \des{\psi}{\kk+\qq, b} \right] \left[ \sum_{c,d} (T^i)_{cd} \cre{\psi}{\kk', c} \des{\psi}{\kk'-\qq, d} \right].
\end{align}
Here $T^i$ are the matrices $\gamma_i$ defined above.  Their orthogonality,
$\Tr(T^iT^j)=4\delta_{ij}$, gives
\begin{align} 
g_i \approx \frac{1}{16} \sum_{a,b,c,d} \mathcal{V}^{(0)}_{ab;cd} (T^i)^*_{ab} (T^i)^*_{cd}.
\end{align}
Thus the same contraction includes both intravalley forward scattering and
intervalley transfer.  It produces the local density-bilinear couplings before
fermionic antisymmetrization; Appendix~\ref{sec:app:finite_doping_bse} constructs
the antisymmetrized vertex used in the PP and PH kernels.

At the QBT anchors, this numerical projection gives nonzero $g_I$, $g_{13}$,
and $g_{15}$; the other components vanish in the bare benchmark.  Although the
screened Coulomb interaction is scalar in the microscopic orbital space, its
band projection contains QBT-spinor form factors.  Their anchor values lie in
the $\sigma_0$ sector.  Intravalley forward scattering contributes mainly to
$g_I\propto\tau_0\sigma_0$, whereas $g_{13}$ and $g_{15}$, proportional to
$\tau_y\sigma_0$ and $\tau_x\sigma_0$, describe intervalley transfer at
$\mathbf K_\Gamma-\mathbf K_M$.  The finite-doping BSE treats these three
couplings as fixed inputs while retaining the exact momentum dependence of the
band energies and Bloch eigenvectors.

The bare benchmark initializes only these three couplings, but one-loop corrections generate terms involving $\sigma_{x,y,z}$. The closed RG equations therefore require the following six-dimensional interaction basis:
\begin{equation}
    \begin{aligned}
    S_{\rm int}&=\frac{1}{2} \int d\tau d^{2} \mathbf{r}\Big[g_{0}\left(\psi^{\dagger}\gamma_{0}\psi\right)^2+g_{2}\left(\psi^{\dagger}\gamma_{2}\psi\right)^2+g_{I}\left(\psi^{\dagger}\psi\right)^2+g_{3}\left(\psi^{\dagger}\gamma_{3}\psi\right)^2+g_{13}\left(\psi^{\dagger}\gamma_{13}\psi\right)^2+g_{15}\left(\psi^{\dagger}\gamma_{15}\psi\right)^2\Big].
    \end{aligned}
\end{equation}

\section{Derivation of the RG equations}

We derive the one-loop renormalization group (RG) equations for the symmetry-allowed four-fermion interactions of the TBCB model. Near each quadratic band touching (QBT), $C_{4z}$ symmetry fixes the non-interacting low-energy Hamiltonian to be~\cite{PhysRevResearch.4.043151}
\begin{equation}
    \mathcal{H}_{0}(\mathbf{k}) = 2t_{\text{eff}} k_x k_y \sigma_x + t_{\text{eff}}(k_x^2 - k_y^2) \sigma_z,\label{eq:h_eff}
\end{equation}
where $t_{\text{eff}}=\frac{1-12\alpha^2}{1+4\alpha^2}t'$ is the effective mass parameter. The corresponding Matsubara Green's function is
\begin{equation}
G_{\mathbf{k}}(i\omega) = \frac{1}{-i\omega + \mathcal{H}_{0}(\mathbf{k})} = \frac{i\omega + \mathcal{H}_{0}(\mathbf{k})}{\omega^2 + \epsilon_{\mathbf{k}}^2},
\label{eq:green_function}
\end{equation}
where the energy dispersion is $\epsilon_{\mathbf{k}} = t_{\text{eff}}(k_x^2+k_y^2)$.

We obtain the coupling flows by integrating out fast fermionic modes in the infinitesimal momentum shell $\Lambda/s < |\mathbf{k}| < \Lambda$. The one-loop vertex corrections are built from the tensor product of two fast Green's functions, with the shell integral
\begin{equation}
\mathcal{I}_\omega = \int_{-\infty}^{\infty} \frac{d \omega}{2 \pi} \int_{\Lambda / s}^{\Lambda} \frac{d^{2} \mathbf{k}}{(2 \pi)^{2}} G_{\mathbf{k}}(i \omega) \otimes G_{\pm\mathbf{k}}(\pm i \omega).
\label{eq:integral_def}
\end{equation}
Substituting Eq.~\eqref{eq:green_function} into Eq.~\eqref{eq:integral_def} and using $\mathcal{H}_0(-\mathbf{k}) = \mathcal{H}_0(\mathbf{k})$, the terms linear in $\omega$ vanish upon frequency integration. The remaining integrand is
\begin{align}
\mathcal{I}_{\omega} = \int_{-\infty}^{\infty} \frac{d\omega}{2\pi} \frac{\mp \omega^2 \mathbb{I} \otimes \mathbb{I} + \mathcal{H}_0(\mathbf{k}) \otimes \mathcal{H}_0(\mathbf{k})}{(\omega^2 + \epsilon_{\mathbf{k}}^2)^2}.    
\end{align}
Performing the standard frequency integrals,
\begin{align}
\int_{-\infty}^{\infty} \frac{d\omega}{2\pi} \frac{\omega^2}{(\omega^2 + \epsilon_{\mathbf{k}}^2)^2} = \int_{-\infty}^{\infty} \frac{d\omega}{2\pi} \frac{\epsilon_{\mathbf{k}}^2}{(\omega^2 + \epsilon_{\mathbf{k}}^2)^2} = \frac{1}{4\epsilon_{\mathbf{k}}}, 
\end{align}
gives $\mathcal{I}_{\omega} = \frac{1}{4\epsilon_{\mathbf{k}}} \left[ \mp \mathbb{I} \otimes \mathbb{I} + \frac{\mathcal{H}_0(\mathbf{k}) \otimes \mathcal{H}_0(\mathbf{k})}{\epsilon_{\mathbf{k}}^2} \right]$. The upper sign ($-$) applies to the particle-hole (PH) channel and the lower sign ($+$) to the particle-particle (PP) channel.

For the momentum-shell integral, we write the Hamiltonian tensor product in polar coordinates as
 \begin{align}
 \frac{\mathcal{H}_0(\mathbf{k}) \otimes \mathcal{H}_0(\mathbf{k})}{\epsilon_{\mathbf{k}}^2} = \sin^2(2\theta)\sigma_x \otimes \sigma_x + \cos^2(2\theta)\sigma_z \otimes \sigma_z + \sin(2\theta)\cos(2\theta)(\sigma_x \otimes \sigma_z + \sigma_z \otimes \sigma_x),    
 \end{align}
the angular average over $\theta \in [0, 2\pi)$ removes the cross terms and gives a factor of $1/2$ for the two squared terms. Defining the density-of-states factor $\mathcal{D}_0 = \frac{1}{8\pi t_{\text{eff}}}$, the radial integral gives $\int_{\Lambda/s}^{\Lambda} \frac{k dk}{2\pi} \frac{1}{4\epsilon_{\mathbf{k}}} = \mathcal{D}_0 \ln s$. The common one-loop kernel is therefore
\begin{equation}
\mathcal{I} = \mathcal{D}_0 \ln s \left[ \mp \mathbb{I} \otimes \mathbb{I} + \frac{1}{2} \left( \sigma_x \otimes \sigma_x + \sigma_z \otimes \sigma_z \right) \right]. 
\label{eq:final_identity}
\end{equation}

With this identity analytically established, we proceed to calculate the corrections to the effective action using the cumulant expansion method. For a generic local quartic interaction $\mathcal{O}_{int} = \frac{1}{2} \sum_{S,T} g_{ST} (\psi^\dagger S \psi)(\psi^\dagger T \psi)$, integrating out the fast modes $\psi_>$ yields
\begin{equation}
\langle e^{-S_{int}} \rangle_> \approx \exp\left( -\langle S_{int} \rangle_> + \frac{1}{2} \left[ \langle S_{int}^2 \rangle_> - \langle S_{int} \rangle_>^2 \right] \right).
\end{equation}
The quadratic cumulant $\frac{1}{2}\langle S_{int}^2 \rangle_>$ generates the one-loop flow. Its contractions fall into three classes: RPA (bubble), vertex, and ladder diagrams. We use $\int_x \equiv \int d\tau d^2r$ below.

The RPA (bubble) contributions arise from the contraction of internal fermion lines within a single interaction loop. For the symmetry-allowed matrix basis $\mathcal{G}$, it is
\begin{equation} 
\Delta S_{eff}^{(RPA)} = \frac{1}{2}\sum_{S, U \in \mathcal{G}} g_{S}g_{U} \int_x \, (\psi^\dagger S \psi) \int \frac{d\omega}{2\pi} \int_{\Lambda/s}^{\Lambda} \frac{d^2k}{(2\pi)^2} \operatorname{Tr}\left[ S G_{\mathbf{k}}(i\omega) U G_{\mathbf{k}}(i\omega) \right] (\psi^\dagger U \psi). \label{eq:RPA_standard}
\end{equation}
Using the Eq.~\eqref{eq:final_identity}, Eq.~\eqref{eq:RPA_standard} becomes
\begin{equation}
\Delta S_{eff}^{(RPA)} = \frac{1}{2} \mathcal{D}_0 \ln s \sum_{S, U \in \mathcal{G}} g_{S}g_{U} \int_x \psi^\dagger S \psi \operatorname{Tr}\left[ -SU + \frac{1}{2}\left(S\sigma_x U \sigma_x +S\sigma_z U \sigma_z\right)\right] \psi^\dagger U \psi.
\label{eq:RPA_explicit}
\end{equation}
The matrix traces give
\begin{equation} 
\Delta S_{eff}^{(RPA)} = - 2 \mathcal{D}_0 \ln s \int_x \left[ g_{0}^2 \left(\psi^{\dagger}\gamma_{0}\psi\right)^2+g_{2}^2 \left(\psi^{\dagger}\gamma_{2}\psi\right)^2 + 2g_{3}^2 \left(\psi^{\dagger}\gamma_{3}\psi\right)^2 \right]. 
\end{equation}

The vertex corrections arise from diagrams where an interaction line bridges a fermion propagator. The general form for these corrections is: 
\begin{equation} 
\Delta S_{eff}^{(V)} = - \sum_{S, U \in \mathcal{G}} g_{S}g_{U} \int_x \, \psi^\dagger S \left( \int \frac{d\omega}{2\pi} \int_{\Lambda/s}^{\Lambda} \frac{d^2k}{(2\pi)^2} G_{\mathbf{k}}(i\omega) U G_{\mathbf{k}}(i\omega) \right) S \psi (\psi^\dagger U \psi). \label{eq:Vertex_standard} 
\end{equation}
Equation~\eqref{eq:final_identity} reduces its matrix structure to
\begin{equation} 
\Delta S_{eff}^{(V)} = - \mathcal{D}_0 \ln s \sum_{S, U \in \mathcal{G}} g_{S}g_{U} \int_x \psi^\dagger S \left[ -U + \frac{1}{2}\left(\sigma_z U \sigma_z+\sigma_x U \sigma_x\right) \right] S \psi (\psi^\dagger U \psi). \label{eq:Vertex_explicit} 
\end{equation}
Substituting the interaction matrices into Eq.~\eqref{eq:Vertex_explicit} gives
\begin{align}
\Delta S_{eff}^{(V)} =  \mathcal{D}_0 \ln s \int_x& \bigg[   g_0  \Big( g_0  - g_2  + g_I    - g_3 + g_{13}  + g_{15} \Big) \left(\psi^\dagger \gamma_0 \psi\right)^2 \nonumber\\
+&g_2  \Big( -g_0  + g_2  + g_I    - g_3 + g_{13}  + g_{15} \Big) \left(\psi^\dagger \gamma_2 \psi\right)^2 \nonumber\\
+& 2g_3  \Big( -g_0  - g_2  + g_I + g_3  - g_{13}  + g_{15} \Big) \left(\psi^\dagger \gamma_3 \psi\right)^2\bigg]. \label{eq:Vertex_Result} 
\end{align} 

The ladder contribution $\Delta S_{eff}^{(L)}$ describes scatterings between interaction vertices and combines the crossed particle-hole (exchange) and particle-particle (pairing) diagrams. We decompose the loop kernel in Eq.~\eqref{eq:final_identity} into its scalar ($\mathbb{I}$) and vector ($\sigma_{x,z}$) parts.

The scalar part ($\mathbb{I} \otimes \mathbb{I}$) enters with opposite signs in the exchange and pairing diagrams. Symmetrizing the sum over $S$ and $U$ combines these terms into
\begin{align}
\text{Scalar Contrib.} &\propto \sum_{S,U} g_S g_U \left[ (\psi^\dagger S U \psi)^2 - (\psi^\dagger S U \psi)(\psi^\dagger U S \psi) \right] \nonumber \\
&= \frac{1}{2} \sum_{S,U} g_S g_U \left[ (\psi^\dagger S U \psi) - (\psi^\dagger U S \psi) \right]^2 \nonumber \\
&= \frac{1}{2} \sum_{S,U} g_S g_U \left[ \psi^\dagger [S, U] \psi \right]^2.
\end{align}
The vector part ($\sigma_x \otimes \sigma_x+\sigma_z \otimes \sigma_z$) has the same sign in both diagrams and gives
\begin{align}
\text{Vector Contrib.} &\propto\frac{1}{2} \sum_{i=x,z}\sum_{S,U} g_S g_U \left[ (\psi^\dagger S \sigma_i U \psi)^2 + (\psi^\dagger S \sigma_i U \psi)(\psi^\dagger U \sigma_i S \psi) \right] \nonumber \\
&= \frac{1}{2} \sum_{S,U} g_S g_U \left[ (\psi^\dagger S \sigma_i U \psi) + (\psi^\dagger U \sigma_i S \psi) \right]^2 \nonumber \\
&= \frac{1}{2} \sum_{S,U} g_S g_U \left[ \psi^\dagger (S \sigma_i U + U \sigma_i S) \psi \right]^2.
\end{align}
Combining the scalar and vector parts and restoring the loop prefactors yields
\begin{align}
\Delta S_{eff}^{(L)} = -\frac{1}{4} \mathcal{D}_0 \ln s \sum_{S,U \in \mathcal{G}} g_{S}g_{U} \int_x \left\{ \left[ \psi^\dagger [S, U] \psi \right]^2 + \frac{1}{2} \sum_{i=x,z}\left[ \psi^\dagger (S \sigma_i U + U \sigma_i S) \psi \right]^2 \right\}.
\label{eq:Ladder_Corrected}
\end{align}

Using the Fierz identities to project the generated four-fermion terms back onto the six-dimensional symmetry-allowed coupling basis gives
\begin{equation}
\begin{aligned}
\Delta S_{eff}^{(L)} = \mathcal{D}_0 \ln s \int_x \bigg\{ 
& \left(\psi^\dagger \gamma_0 \psi\right)^2 \Big[ -\frac{1}{2}(g_0^2 + g_2^2 + g_I^2 + g_3^2 + g_{13}^2 + g_{15}^2) - g_0 g_2 - 2g_2 g_3 + g_I(g_{13} + g_{15}) + g_3 g_{13} - 2g_{13} g_{15} \Big] \\
+ & \left(\psi^\dagger \gamma_2 \psi\right)^2 \Big[ -\frac{1}{2}(g_0^2 + g_2^2 + g_I^2 + g_3^2 + g_{13}^2 + g_{15}^2) - g_0(g_2 + 2g_3) + g_I(g_{13} + g_{15}) + g_3 g_{13} - 2g_{13} g_{15} \Big] \\
+ & \left(\psi^\dagger \psi\right)^2 \Big[ -2g_0(g_2 + g_3) - g_I(g_0 + g_2) - 2g_2 g_3 + 2g_I(g_{13} + g_{15}) + 2g_{13}(g_3 - g_{15}) \Big] \\
+ & \left(\psi^\dagger \gamma_3 \psi\right)^2 \Big[ -2g_0 g_2 - g_3(g_0 + g_2) + 2g_{13}(g_3 + g_{15}) \Big] \\
+ & \left(\psi^\dagger \gamma_{13} \psi\right)^2 \Big[ -g_{13}(g_0 + g_2 - 2g_I + 2g_{15}) \Big] \\
+ & \left(\psi^\dagger \gamma_{15} \psi\right)^2 \Big[ -2g_0(g_2 + g_3) - g_{15}(g_0 + g_2 - 2g_I) - 2g_2 g_3 + 2g_3 g_{13} \Big]
\bigg\}.
\end{aligned}
\label{eq:Ladder_Result}
\end{equation}

The total one-loop correction is $\Delta S_{eff} = \Delta S_{eff}^{(RPA)} + \Delta S_{eff}^{(V)} + \Delta S_{eff}^{(L)}$. Matching it to $S_{int} = \frac{1}{2} \int_x \sum_i g_i (\psi^\dagger \gamma_i \psi)^2$ gives $\frac{1}{2} \Delta g_i = \mathcal{D}_0 \ln s \cdot \beta_i^{(0)}$, where $\beta_i^{(0)}$ is the polynomial coefficients obtained from the loop integrals. Keeping the density-of-states factor $\mathcal{D}_0$ explicit, the six coupling flows are
\begin{align}
\frac{dg_0}{d\ln s}
&=2\mathcal D_0\bigg[
-\frac12\left(3g_0^2+g_2^2+g_I^2+g_3^2+g_{13}^2+g_{15}^2\right)
+g_0\left(g_I-g_3+g_{13}+g_{15}-2g_2\right)
\nonumber\\
&\hspace{3.5cm}
-2g_2g_3+g_I(g_{13}+g_{15})+g_3g_{13}-2g_{13}g_{15}
\bigg],\\
\frac{dg_2}{d\ln s}
&=2\mathcal D_0\bigg[
-\frac12\left(g_0^2+3g_2^2+g_I^2+g_3^2+g_{13}^2+g_{15}^2\right)
+g_2\left(g_I-g_3+g_{13}+g_{15}-2g_0\right)
\nonumber\\
&\hspace{3.5cm}
-2g_0g_3+g_I(g_{13}+g_{15})+g_3g_{13}-2g_{13}g_{15}
\bigg],\\
\frac{dg_I}{d\ln s}
&=2\mathcal D_0\bigg[
-g_I(g_0+g_2-2g_{13}-2g_{15})
-2(g_0g_2+g_0g_3+g_2g_3)
+2g_{13}(g_3-g_{15})
\bigg],\\
\frac{dg_3}{d\ln s}
&=2\mathcal D_0\bigg[
-2g_3^2+g_3(2g_I-3g_0-3g_2+2g_{15})
-2g_0g_2+2g_{13}g_{15}
\bigg],\\
\frac{dg_{13}}{d\ln s}
&=2\mathcal D_0\left[-g_{13}(g_0+g_2-2g_I+2g_{15})\right],\\
\frac{dg_{15}}{d\ln s}
&=2\mathcal D_0\bigg[
-g_{15}(g_0+g_2-2g_I)
-2(g_0g_2+g_0g_3+g_2g_3)+2g_3g_{13}
\bigg].
\end{align}

\section{Susceptibility and source maps}

We diagnose competing weak-coupling ordering tendencies from the renormalization of symmetry-breaking source fields.

\begin{figure*}[htbp]
    \centering 
    
    \begin{minipage}{0.30\linewidth} 
        \centering
        \begin{tikzpicture}[baseline=(current bounding box.center)]
            \begin{feynman}
                \vertex (in);
                \vertex [right=1.2cm of in] (m);
                \vertex [right=1.2cm of m] (out);
                \vertex [above=1.0cm of m] (b_bot);
                \vertex [above=1.2cm of b_bot] (b_top);
                \vertex [above=0.8cm of b_top] (top);

                \diagram* {
                    (in) -- [fermion] (m) -- [fermion] (out),
                    (b_bot) -- [photon] (m),
                    (b_bot) -- [fermion, half left] (b_top),
                    (b_top) -- [fermion, half left] (b_bot),
                    (top) -- [photon] (b_top),
                };
            \end{feynman}
        \end{tikzpicture}
    \end{minipage}
    \hspace{-0.5cm} 
    \begin{minipage}{0.30\linewidth}
        \centering
        \begin{tikzpicture}[baseline=(current bounding box.center)]
            \begin{feynman}
                \vertex (l);
                \vertex [right=1.2cm of l] (c);
                \vertex [right=1.2cm of c] (r_dummy);
                \vertex [above=0.8cm of r_dummy] (t);
                \vertex [below=0.8cm of r_dummy] (b);
                \vertex [right=0.8cm of t] (out_t);
                \vertex [right=0.8cm of b] (out_b);

                \diagram* {
                    (l) -- [photon] (c),
                    (c) -- [fermion] (t),
                    (c) -- [anti fermion] (b),
                    (t) -- [photon] (b),
                    (t) -- [fermion] (out_t),
                    (b) -- [anti fermion] (out_b),
                };
            \end{feynman}
        \end{tikzpicture}
    \end{minipage}
    \hspace{-0.5cm} 
    \begin{minipage}{0.30\linewidth}
        \centering
        \begin{tikzpicture}[baseline=(current bounding box.center)]
            \begin{feynman}
                \vertex (l);
                \vertex [right=1.2cm of l] (c);
                \vertex [right=1.2cm of c] (r_dummy);
                \vertex [above=0.8cm of r_dummy] (t);
                \vertex [below=0.8cm of r_dummy] (b);
                \vertex [right=0.8cm of t] (out_t);
                \vertex [right=0.8cm of b] (out_b);

                \diagram* {
                    (l) -- [photon] (c),
                    (c) -- [anti fermion] (t),
                    (c) -- [anti fermion] (b), 
                    (t) -- [photon] (b),
                    (t) -- [anti fermion] (out_t),
                    (b) -- [anti fermion] (out_b), 
                };
            \end{feynman}
        \end{tikzpicture}
    \end{minipage}
    
    \caption{One-loop diagrams that renormalize the source coupling $\Delta_{\mathcal{O}}$. The left and middle diagrams give the bubble term $\Pi_{\mathcal{O}M}$ and vertex term $\Upsilon_{\mathcal{O}M}$ in the particle-hole flow, Eq.~\eqref{equ:phchannel}. The right diagram gives the particle-particle vertex correction in Eq.~\eqref{equ:ppchannel}.}
    \label{app:fig:sus_diagram}
\end{figure*}
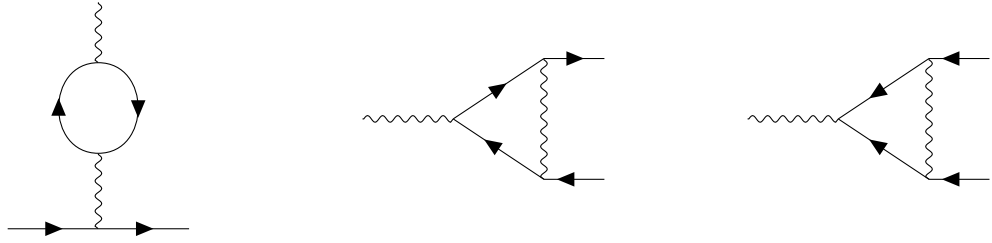

For a generic particle-hole source $\Delta_{m}^{ph}\psi^{\dagger}\mathcal{O}_{m}\psi$, the spinless problem has 16 orbital matrices. In the spin-$1/2$ problem, they generate 16 charge multiplets, $\mathcal O_m\otimes s_0$, and 16 spin-vector multiplets, $\mathcal O_m\otimes\vec s$. Each spin-vector row represents three $SU(2)$-degenerate components. The 32 labels below therefore count source multiplets, not individual matrices in the 64-dimensional $U(8)$ basis. The renormalized spinless source is
\begin{equation}
\begin{aligned}
\Delta_{m}^{ph}(s) \psi_{<}^{\dagger} \mathcal{O}_{m} \psi_{<}(\mathbf{r}, \tau) &= s^{2} \Delta_{m}^{ph}(1) \psi_{<}^{\dagger} \mathcal{O}_{m} \psi_{<} \\
&+ s^{2} \Delta_{m}^{ph}(1) \sum_{MN} g_{M N} \Big[\Pi_{\mathcal{O}_{m}M} \psi_{<}^{\dagger} N \psi_{<}(\mathbf{r}, \tau)+ \Pi_{\mathcal{O}_{m}N} \psi_{<}^{\dagger} M \psi_{<}(\mathbf{r}, \tau)\Big]\\
&- s^{2} \Delta_{m}^{ph}(1) \sum_{MN} g_{M N} \Big[\psi_{<}^{\dagger} \Upsilon_{\mathcal{O}_{m} M N} \psi_{<}(\mathbf{r}, \tau)+\psi_{<}^{\dagger} \Upsilon_{\mathcal{O}_{m} N M} \psi_{<}(\mathbf{r}, \tau)\Big],
\end{aligned}\label{equ:phchannel}
\end{equation}
where $g_{MN}$ denotes an independent interaction coupling. For a diagonal interaction $g_M (\psi^\dagger M \psi)^2$, the sum is restricted to $M=N$. The one-loop matrix elements, corresponding respectively to the closed-loop diagram $\Pi$ and the vertex correction diagram $\Upsilon$ (see Fig.~\ref{app:fig:sus_diagram}), are explicitly given by:
\begin{equation}
\Pi_{\mathcal{O}_{m}M}=\int_{-\infty}^{\infty} \frac{d \omega}{2 \pi} \int_{\Lambda / s}^{\Lambda} \frac{d^{2} \mathbf{k}}{(2 \pi)^{2}} \operatorname{Tr}\left[G_{\mathbf{k}}(i \omega) \mathcal{O}_{m} G_{\mathbf{k}}(i \omega) M\right],
\end{equation}
\begin{equation}
\Upsilon_{\mathcal{O}_{m}MN}=\int_{-\infty}^{\infty} \frac{d \omega}{2 \pi} \int_{\Lambda / s}^{\Lambda} \frac{d^{2} \mathbf{k}}{(2 \pi)^{2}} M G_{\mathbf{k}}(i \omega) \mathcal{O}_{m} G_{\mathbf{k}}(i \omega) N.
\end{equation}
Trace orthogonality makes $\Pi_{\mathcal{O}_{m}M}$ nonzero only for $\mathcal{O}_{m}=M$. For $X=ph$ or $pp$, the source flow has the general index structure
\begin{equation}
    \frac{d\Delta_i^X}{d\ln s}
    =2\Delta_i^X+\mathcal D_0\sum_{j,a}
    \mathcal F^X_{ij;a}g_a\Delta_j^X,
    \label{undig}
\end{equation}
where $i,j$ label sources, $a$ labels the interaction couplings, and $\mathcal F^X_{ij;a}$ includes the diagrammatic signs. The spatial parities in Table~\ref{tab:symmetry_classification}, together with the $SU(2)$ representation in the spinful problem, forbid mixing between distinct source multiplets. In the source bases used here, $\mathcal F^X_{ij;a}=\delta_{ij}M^X_{ia}$, so
\begin{equation}
    \frac{d\ln\Delta_i^X}{d\ln s}
    =2+\mathcal D_0\sum_a M^X_{ia}g_a.
    \label{eq:decoupled_source_flow}
\end{equation}

For the particle-particle channel, the source $\Delta_{n}^{pp}\psi^T\mathcal{O}_{n}\psi$ contains two annihilation operators, so Fermi statistics require $\mathcal{O}_{n}^{T}=-\mathcal{O}_{n}$. Six matrices in the $U(4)$ basis satisfy this condition: $\{\gamma_1,\gamma_3,\gamma_{13},\gamma_{02},\gamma_{05},\gamma_{25}\}$. Their one-loop flow is
\begin{equation}
\begin{aligned}
\Delta^{pp}_{n}(s) \psi_{a}^{<} \mathcal{O}_{a b}^{(n)} \psi_{b}^{<}&= s^{2} \Delta^{pp}_{n}(1) \psi_{a}^{<} \mathcal{O}_{a b}^{(n)} \psi_{b}^{<} \\
&+ s^{2} \Delta^{pp}_{n}(1) \sum_{M N} g_{M N} \int \frac{d \omega}{2 \pi} \int_{\Lambda / s}^{\Lambda} \frac{d^{2} \mathbf{k}}{(2 \pi)^{2}} \Big[ \\
&\quad \mathcal{O}_{\alpha \beta}^{(n)} G_{\mathbf{k}, \beta \gamma}(i \omega) M_{\gamma \sigma} \psi_{\sigma<}(\mathbf{r}, \tau) G_{-\mathbf{k}, \alpha \delta}(-i \omega) N_{\delta \nu} \psi_{\nu<}(\mathbf{r}, \tau) \\
&\quad + \mathcal{O}_{\alpha \beta}^{(n)} G_{\mathbf{k}, \beta \gamma}(i \omega) N_{\gamma \sigma} \psi_{\sigma<}(\mathbf{r}, \tau) G_{-\mathbf{k}, \alpha \delta}(-i \omega) M_{\delta \nu} \psi_{\nu<}(\mathbf{r}, \tau) \Big].
\end{aligned}
\label{equ:ppchannel}
\end{equation}
The integral term is the particle-particle ladder diagram in Fig.~\ref{app:fig:sus_diagram}. We identify the leading one-loop source by integrating Eq.~\eqref{eq:decoupled_source_flow} together with the coupling flow.

For the spinless six-coupling TBCB theory, the sources with nonzero one-loop vertex corrections obey
\begin{align}
\frac{d \ln \Delta_{0}^{ph}}{d \ln s} &= 2 - (3g_{0} + g_{2} - g_{I} + g_{3} - g_{13} - g_{15})\mathcal{D}_0, \\
\frac{d \ln \Delta_{1}^{ph}}{d \ln s} &= 2 - (2g_{0} + 2g_{2} - 2g_{I} + 2g_{3} + 2g_{13} + 2g_{15})\mathcal{D}_0, \\
\frac{d \ln \Delta_{2}^{ph}}{d \ln s} &= 2 - (g_{0} + 3g_{2} - g_{I} + g_{3} - g_{13} - g_{15})\mathcal{D}_0, \\
\frac{d \ln \Delta_{3}^{ph}}{d \ln s} &= 2 - (2g_{0} + 2g_{2} - 2g_{I} + 6g_{3} + 2g_{13} - 2g_{15})\mathcal{D}_0, \\
\frac{d \ln \Delta_{5}^{ph}}{d \ln s} &= 2 - (2g_{0} + 2g_{2} - 2g_{I} + 2g_{3} - 2g_{13} + 2g_{15})\mathcal{D}_0, \\
\frac{d \ln \Delta_{01}^{ph}}{d \ln s} &= 2 - (g_{0} - g_{2} - g_{I} - g_{3} + g_{13} + g_{15})\mathcal{D}_0, \\
\frac{d \ln \Delta_{02}^{ph}}{d \ln s} &= 2 - (2g_{0} + 2g_{2} - 2g_{I} - 2g_{3} - 2g_{13} - 2g_{15})\mathcal{D}_0, \\
\frac{d \ln \Delta_{03}^{ph}}{d \ln s} &= 2 - (g_{0} - g_{2} - g_{I} + g_{3} + g_{13} - g_{15})\mathcal{D}_0, \\
\frac{d \ln \Delta_{05}^{ph}}{d \ln s} &= 2 - (g_{0} - g_{2} - g_{I} - g_{3} - g_{13} + g_{15})\mathcal{D}_0, \\
\frac{d \ln \Delta_{12}^{ph}}{d \ln s} &= 2 + (g_{0} - g_{2} + g_{I} + g_{3} - g_{13} - g_{15})\mathcal{D}_0, \\
\frac{d \ln \Delta_{23}^{ph}}{d \ln s} &= 2 + (g_{0} - g_{2} + g_{I} - g_{3} - g_{13} + g_{15})\mathcal{D}_0, \\
\frac{d \ln \Delta_{25}^{ph}}{d \ln s} &= 2 + (g_{0} - g_{2} + g_{I} + g_{3} + g_{13} - g_{15})\mathcal{D}_0, \\
\frac{d \ln \Delta_{05}^{pp}}{d \ln s} &= 2 + (g_{0} - g_{2} - g_{I} + g_{3} + g_{13} + g_{15})\mathcal{D}_0, \\
\frac{d \ln \Delta_{13}^{pp}}{d \ln s} &= 2 - (2g_{0} + 2g_{2} + 2g_{I} + 2g_{3} - 2g_{13} - 2g_{15})\mathcal{D}_0, \\
\frac{d \ln \Delta_{25}^{pp}}{d \ln s} &= 2 - (g_{0} - g_{2} + g_{I} - g_{3} - g_{13} - g_{15})\mathcal{D}_0.
\end{align}
The PH sources $\mathbb I$, $\gamma_{13}$, $\gamma_{15}$, and $\gamma_{35}$ and the PP sources $\gamma_1$, $\gamma_3$, and $\gamma_{02}$ have zero one-loop vertex correction in this basis and therefore retain only the common tree-level term $2$.

Writing $\ell=\ln s$, the one-loop couplings formally run away at a scale
$\ell_c$ defined by
\begin{equation}
    \lim_{\ell\to\ell_c^-}
    \max_a |\mathcal D_0 g_a(\ell)|=\infty .
    \label{eq:rg_runaway_scale}
\end{equation}
Since the one-loop expansion ceases to be controlled before this formal
divergence is reached, we stop all coupling and source flows at the same
finite scale $\ell_\star$, defined by
$\|\mathcal D_0\mathbf g(\ell_\star)\|_2=1$.  This common stopping scale
allows different source channels to be compared within the same perturbative
RG trajectory.  Since the QBT has dynamic exponent $z=2$, the corresponding
running energy scale is
$E_\star\sim t_{\mathrm{eff}}\Lambda^2e^{-2\ell_\star}$.

Following source-field RG analyses of quadratic band-touching
systems~\cite{PhysRevB.82.205106,PhysRevB.86.075467}, we use the finite-scale renormalization of symmetry-breaking sources as a diagnostic of competing ordering tendencies.  All symmetry-allowed sources are initialized with the same infinitesimal normalization and evolved together with the couplings to $\ell_\star$.  Integrating Eq.~\eqref{eq:decoupled_source_flow}, we define the interaction-induced RG enhancement
\begin{equation}
 H_i(\ell_\star)=\int_0^{\ell_\star}d\ell\,
 \mathcal D_0\sum_a M_{ia}g_a(\ell)
 =\ln\frac{\Delta_i(\ell_\star)}{\Delta_i(0)}-2\ell_\star.
 \label{eq:accumulated_source_selector}
\end{equation}
Thus \(H_i\) is the integrated interaction-induced correction to the source scaling, with the common tree-level contribution removed. At each point, the source with the largest \(H_i\) is identified as the leading finite-scale RG source. This prescription compares the full perturbative RG flow.

\section{Spin-$1/2$ RG equations and source flows}
For spin-$1/2$ fermions, $\psi = (\psi_{\uparrow}, \psi_{\downarrow})^T$ and the internal matrix basis expands from $U(4)$ to $U(8)$. In the absence of spin-orbit coupling, global $SU(2)$ spin rotation symmetry splits each orbital interaction in Eq.~\eqref{equ:15interactions} into charge and spin channels. The most general symmetry-allowed interaction is
$$ S_{\rm int}^{(s=1/2)} = \frac{1}{2} \int d\tau d^2r \sum_{i=1}^{16} \left[ g_{i}^{c} (\psi^\dagger \gamma_i s_0 \psi)^2 + g_{i}^{s} (\psi^\dagger \gamma_i \vec{s} \psi) \cdot (\psi^\dagger \gamma_i \vec{s} \psi) \right], $$
where $s_0$ is the $2 \times 2$ identity in spin space and $\vec{s} = (s_x, s_y, s_z)$ are the spin Pauli matrices. The point-group and nonsymmorphic symmetries discussed in Appendix B forbid mixed orbital terms $(\psi^\dagger\gamma_i\psi)(\psi^\dagger\gamma_j\psi)$ with $i\neq j$. The lattice symmetries therefore allow 32 couplings before imposing Fierz relations.

We remove the redundancy among these couplings with the generalized $SU(8)$ Fierz identity. For arbitrary $8 \times 8$ Hermitian matrices $M$ and $N$,
$$ (\psi^\dagger M \psi)(\psi^\dagger N \psi) = -\frac{1}{64} \sum_{A,B=1}^{64} \text{Tr}[M \Lambda_A N \Lambda_B] (\psi^\dagger \Lambda_B \psi)(\psi^\dagger \Lambda_A \psi), $$
where $\{\Lambda_A\} = \{\gamma_i \otimes s_0, \gamma_i \otimes \vec{s}\}$ is the complete 64-dimensional $U(8)$ matrix basis.

Compared with Eq.~\eqref{eq:fierz_spinless}, the spinful identity uses a $64\times64$ trace algebra and the prefactor $-1/64$. Applying it to $\mathcal V=(g_1^c,\ldots,g_{16}^c,g_1^s,\ldots)$ gives $\mathcal F\mathcal V=0$. The null space leaves 16 independent couplings, for which we choose the charge-channel representatives $g_i^c$; the spin couplings $g_i^s$ are then fixed by the Fierz relations. Their one-loop flows are
\begin{equation}
\begin{split}
\frac{dg_{0}}{d\ln s} = \mathcal{D}_0 \big[ & - g_{I}^{2} + 2 g_{I} g_{0} - 7 g_{0}^{2} - 2 g_{0} g_{01} - 2 g_{0} g_{02} - 2 g_{0} g_{03} - 2 g_{0} g_{05} - 2 g_{0} g_{1} + 2 g_{0} g_{12} \\
& + 2 g_{0} g_{13} + 2 g_{0} g_{15} - 4 g_{0} g_{2} + 2 g_{0} g_{23} + 2 g_{0} g_{25} - 2 g_{0} g_{3} + 2 g_{0} g_{35} - 2 g_{0} g_{5} \\
& - g_{01}^{2} + 4 g_{01} g_{1} - 2 g_{01} g_{12} - g_{02}^{2} + 4 g_{02} g_{2} - g_{03}^{2} - 2 g_{03} g_{23} + 4 g_{03} g_{3} \\
& - g_{05}^{2} - 2 g_{05} g_{25} + 4 g_{05} g_{5} - g_{1}^{2} - g_{12}^{2} - g_{13}^{2} - g_{15}^{2} - g_{2}^{2} - g_{23}^{2} - g_{25}^{2} - g_{3}^{2} - g_{35}^{2} - g_{5}^{2} \big]
\end{split}
\end{equation}

\begin{equation}
\begin{split}
\frac{dg_{01}}{d\ln s} = \mathcal{D}_0 \big[ & 2 g_{I} g_{01} - 2 g_{I} g_{35} - 4 g_{0} g_{01} + 4 g_{0} g_{1} - 2 g_{0} g_{12} - 6 g_{01}^{2} - 2 g_{01} g_{02} - 2 g_{01} g_{03} \\
& - 2 g_{01} g_{05} - 2 g_{01} g_{1} - 2 g_{01} g_{12} - 2 g_{01} g_{13} - 2 g_{01} g_{15} + 2 g_{01} g_{23} + 2 g_{01} g_{25} \\
& + 2 g_{01} g_{3} + 2 g_{01} g_{35} + 2 g_{01} g_{5} - 2 g_{02} g_{1} + 4 g_{02} g_{12} + 4 g_{03} g_{13} + 4 g_{05} g_{15} \\
& - 2 g_{12} g_{2} - 2 g_{13} g_{3} - 2 g_{15} g_{5} \big]
\end{split}
\end{equation}

\begin{equation}
\begin{split}
\frac{dg_{2}}{d\ln s} = \mathcal{D}_0 \big[ & - g_{I}^{2} + 2 g_{I} g_{2} - g_{0}^{2} + 4 g_{0} g_{02} - 4 g_{0} g_{2} - g_{01}^{2} - 2 g_{01} g_{12} + 2 g_{01} g_{2} - g_{02}^{2} \\
& - 2 g_{02} g_{2} - g_{03}^{2} + 2 g_{03} g_{2} - 2 g_{03} g_{23} - g_{05}^{2} + 2 g_{05} g_{2} - 2 g_{05} g_{25} - g_{1}^{2} \\
& + 4 g_{1} g_{12} - 2 g_{1} g_{2} - g_{12}^{2} - 2 g_{12} g_{2} - g_{13}^{2} + 2 g_{13} g_{2} - g_{15}^{2} + 2 g_{15} g_{2} - 7 g_{2}^{2} \\
& - 2 g_{2} g_{23} - 2 g_{2} g_{25} - 2 g_{2} g_{3} + 2 g_{2} g_{35} - 2 g_{2} g_{5} - g_{23}^{2} + 4 g_{23} g_{3} - g_{25}^{2} + 4 g_{25} g_{5} \\
& - g_{3}^{2} - g_{35}^{2} - g_{5}^{2} \big]
\end{split}
\end{equation}

\begin{equation}
\begin{split}
\frac{dg_{12}}{d\ln s} = \mathcal{D}_0 \big[ & 2 g_{I} g_{12} - 2 g_{I} g_{35} - 2 g_{0} g_{01} + 4 g_{01} g_{02} - 2 g_{01} g_{12} - 2 g_{01} g_{2} - 2 g_{02} g_{1} \\
& - 2 g_{02} g_{12} + 2 g_{03} g_{12} + 2 g_{05} g_{12} - 2 g_{1} g_{12} + 4 g_{1} g_{2} - 6 g_{12}^{2} - 2 g_{12} g_{13} \\
& - 2 g_{12} g_{15} - 4 g_{12} g_{2} - 2 g_{12} g_{23} - 2 g_{12} g_{25} + 2 g_{12} g_{3} + 2 g_{12} g_{35} + 2 g_{12} g_{5} \\
& + 4 g_{13} g_{23} - 2 g_{13} g_{3} + 4 g_{15} g_{25} - 2 g_{15} g_{5} \big]
\end{split}
\end{equation}

\begin{equation}
\begin{split}
\frac{dg_{1}}{d\ln s} = \mathcal{D}_0 \big[ & 4 g_{I} g_{1} + 4 g_{0} g_{01} - 6 g_{0} g_{1} - 2 g_{01} g_{02} - 4 g_{01} g_{1} + 4 g_{02} g_{1} - 2 g_{02} g_{12} \\
& + 4 g_{03} g_{1} - 2 g_{03} g_{13} + 4 g_{05} g_{1} - 2 g_{05} g_{15} - 12 g_{1}^{2} - 4 g_{1} g_{12} - 4 g_{1} g_{13} \\
& - 4 g_{1} g_{15} - 6 g_{1} g_{2} + 4 g_{1} g_{23} + 4 g_{1} g_{25} - 4 g_{1} g_{3} + 4 g_{1} g_{35} - 4 g_{1} g_{5} \\
& + 4 g_{12} g_{2} - 2 g_{13} g_{23} + 4 g_{13} g_{3} - 2 g_{15} g_{25} + 4 g_{15} g_{5} \big]
\end{split}
\end{equation}

\begin{equation}
\begin{split}
\frac{dg_{02}}{d\ln s} = \mathcal{D}_0 \big[ & 4 g_{I} g_{02} - 6 g_{0} g_{02} + 4 g_{0} g_{2} - 4 g_{01} g_{02} - 2 g_{01} g_{1} + 4 g_{01} g_{12} - 12 g_{02}^{2} \\
& - 4 g_{02} g_{03} - 4 g_{02} g_{05} + 4 g_{02} g_{1} - 4 g_{02} g_{12} + 4 g_{02} g_{13} + 4 g_{02} g_{15} - 6 g_{02} g_{2} \\
& - 4 g_{02} g_{23} - 4 g_{02} g_{25} + 4 g_{02} g_{3} + 4 g_{02} g_{35} + 4 g_{02} g_{5} + 4 g_{03} g_{23} - 2 g_{03} g_{3} \\
& + 4 g_{05} g_{25} - 2 g_{05} g_{5} - 2 g_{1} g_{12} - 2 g_{23} g_{3} - 2 g_{25} g_{5} \big]
\end{split}
\end{equation}

\begin{equation}
\begin{split}
\frac{dg_{I}}{d\ln s} = \mathcal{D}_0 \big[ & - 2 g_{I} g_{0} - 2 g_{I} g_{2} - 2 g_{01} g_{35} - 2 g_{03} g_{15} - 2 g_{05} g_{13} - 2 g_{12} g_{35} - 2 g_{13} g_{25} - 2 g_{15} g_{23} \big]
\end{split}
\end{equation}

\begin{equation}
\begin{split}
\frac{dg_{35}}{d\ln s} = \mathcal{D}_0 \big[ & - 2 g_{I} g_{01} - 2 g_{I} g_{12} - 2 g_{0} g_{35} + 4 g_{03} g_{05} - 2 g_{03} g_{5} - 2 g_{05} g_{3} \\
& + 4 g_{13} g_{15} - 2 g_{2} g_{35} + 4 g_{23} g_{25} - 2 g_{23} g_{5} - 2 g_{25} g_{3} + 4 g_{3} g_{5} \big]
\end{split}
\end{equation}

\begin{equation}
\begin{split}
\frac{dg_{3}}{d\ln s} = \mathcal{D}_0 \big[ & 4 g_{I} g_{3} + 4 g_{0} g_{03} - 6 g_{0} g_{3} - 2 g_{01} g_{13} + 4 g_{01} g_{3} - 2 g_{02} g_{03} - 2 g_{02} g_{23} \\
& + 4 g_{02} g_{3} - 4 g_{03} g_{3} + 4 g_{05} g_{3} - 2 g_{05} g_{35} + 4 g_{1} g_{13} - 4 g_{1} g_{3} - 2 g_{12} g_{13} \\
& + 4 g_{12} g_{3} - 4 g_{13} g_{3} + 4 g_{15} g_{3} + 4 g_{2} g_{23} - 6 g_{2} g_{3} - 4 g_{23} g_{3} + 4 g_{25} g_{3} \\
& - 2 g_{25} g_{35} - 12 g_{3}^{2} - 4 g_{3} g_{35} - 4 g_{3} g_{5} + 4 g_{35} g_{5} \big]
\end{split}
\end{equation}

\begin{equation}
\begin{split}
\frac{dg_{5}}{d\ln s} = \mathcal{D}_0 \big[ & 4 g_{I} g_{5} + 4 g_{0} g_{05} - 6 g_{0} g_{5} - 2 g_{01} g_{15} + 4 g_{01} g_{5} - 2 g_{02} g_{05} - 2 g_{02} g_{25} \\
& + 4 g_{02} g_{5} - 2 g_{03} g_{35} + 4 g_{03} g_{5} - 4 g_{05} g_{5} + 4 g_{1} g_{15} - 4 g_{1} g_{5} - 2 g_{12} g_{15} \\
& + 4 g_{12} g_{5} + 4 g_{13} g_{5} - 4 g_{15} g_{5} + 4 g_{2} g_{25} - 6 g_{2} g_{5} - 2 g_{23} g_{35} + 4 g_{23} g_{5} \\
& - 4 g_{25} g_{5} + 4 g_{3} g_{35} - 4 g_{3} g_{5} - 4 g_{35} g_{5} - 12 g_{5}^{2} \big]
\end{split}
\end{equation}

\begin{equation}
\begin{split}
\frac{dg_{03}}{d\ln s} = \mathcal{D}_0 \big[ & 2 g_{I} g_{03} - 2 g_{I} g_{15} - 4 g_{0} g_{03} - 2 g_{0} g_{23} + 4 g_{0} g_{3} - 2 g_{01} g_{03} + 4 g_{01} g_{13} \\
& - 2 g_{02} g_{03} + 4 g_{02} g_{23} - 2 g_{02} g_{3} - 6 g_{03}^{2} - 2 g_{03} g_{05} + 2 g_{03} g_{1} + 2 g_{03} g_{12} \\
& - 2 g_{03} g_{13} + 2 g_{03} g_{15} - 2 g_{03} g_{23} + 2 g_{03} g_{25} - 2 g_{03} g_{3} - 2 g_{03} g_{35} + 2 g_{03} g_{5} \\
& + 4 g_{05} g_{35} - 2 g_{1} g_{13} - 2 g_{2} g_{23} - 2 g_{35} g_{5} \big]
\end{split}
\end{equation}

\begin{equation}
\begin{split}
\frac{dg_{05}}{d\ln s} = \mathcal{D}_0 \big[ & 2 g_{I} g_{05} - 2 g_{I} g_{13} - 4 g_{0} g_{05} - 2 g_{0} g_{25} + 4 g_{0} g_{5} - 2 g_{01} g_{05} + 4 g_{01} g_{15} \\
& - 2 g_{02} g_{05} + 4 g_{02} g_{25} - 2 g_{02} g_{5} - 2 g_{03} g_{05} + 4 g_{03} g_{35} - 6 g_{05}^{2} + 2 g_{05} g_{1} \\
& + 2 g_{05} g_{12} + 2 g_{05} g_{13} - 2 g_{05} g_{15} + 2 g_{05} g_{23} - 2 g_{05} g_{25} + 2 g_{05} g_{3} \\
& - 2 g_{05} g_{35} - 2 g_{05} g_{5} - 2 g_{1} g_{15} - 2 g_{2} g_{25} - 2 g_{3} g_{35} \big]
\end{split}
\end{equation}

\begin{equation}
\begin{split}
\frac{dg_{13}}{d\ln s} = \mathcal{D}_0 \big[ & - 2 g_{I} g_{05} - 2 g_{I} g_{25} - 2 g_{0} g_{13} + 4 g_{01} g_{03} - 2 g_{01} g_{3} - 2 g_{03} g_{1} \\
& - 2 g_{1} g_{23} + 4 g_{1} g_{3} + 4 g_{12} g_{23} - 2 g_{12} g_{3} - 2 g_{13} g_{2} + 4 g_{15} g_{35} \big]
\end{split}
\end{equation}

\begin{equation}
\begin{split}
\frac{dg_{15}}{d\ln s} = \mathcal{D}_0 \big[ & - 2 g_{I} g_{03} - 2 g_{I} g_{23} - 2 g_{0} g_{15} + 4 g_{01} g_{05} - 2 g_{01} g_{5} - 2 g_{05} g_{1} \\
& - 2 g_{1} g_{25} + 4 g_{1} g_{5} + 4 g_{12} g_{25} - 2 g_{12} g_{5} + 4 g_{13} g_{35} - 2 g_{15} g_{2} \big]
\end{split}
\end{equation}

\begin{equation}
\begin{split}
\frac{dg_{23}}{d\ln s} = \mathcal{D}_0 \big[ & - 2 g_{I} g_{15} + 2 g_{I} g_{23} - 2 g_{0} g_{03} + 2 g_{01} g_{23} + 4 g_{02} g_{03} - 2 g_{02} g_{23} \\
& - 2 g_{02} g_{3} - 2 g_{03} g_{2} - 2 g_{03} g_{23} + 2 g_{05} g_{23} - 2 g_{1} g_{13} + 2 g_{1} g_{23} \\
& + 4 g_{12} g_{13} - 2 g_{12} g_{23} - 2 g_{13} g_{23} + 2 g_{15} g_{23} - 4 g_{2} g_{23} + 4 g_{2} g_{3} \\
& - 6 g_{23}^{2} - 2 g_{23} g_{25} - 2 g_{23} g_{3} - 2 g_{23} g_{35} + 2 g_{23} g_{5} + 4 g_{25} g_{35} - 2 g_{35} g_{5} \big]
\end{split}
\end{equation}

\begin{equation}
\begin{split}
\frac{dg_{25}}{d\ln s} = \mathcal{D}_0 \big[ & - 2 g_{I} g_{13} + 2 g_{I} g_{25} - 2 g_{0} g_{05} + 2 g_{01} g_{25} + 4 g_{02} g_{05} - 2 g_{02} g_{25} \\
& - 2 g_{02} g_{5} + 2 g_{03} g_{25} - 2 g_{05} g_{2} - 2 g_{05} g_{25} - 2 g_{1} g_{15} + 2 g_{1} g_{25} \\
& + 4 g_{12} g_{15} - 2 g_{12} g_{25} + 2 g_{13} g_{25} - 2 g_{15} g_{25} - 4 g_{2} g_{25} + 4 g_{2} g_{5} \\
& - 2 g_{23} g_{25} + 4 g_{23} g_{35} - 6 g_{25}^{2} + 2 g_{25} g_{3} - 2 g_{25} g_{35} - 2 g_{25} g_{5} - 2 g_{3} g_{35} \big]
\end{split}
\end{equation}

We order the 16 independent couplings as
\begin{equation}
\begin{aligned}
\mathbf g^c=(&g_0,g_{01},g_2,g_{12},g_1,g_{02},g_I,g_{35},g_3,g_5,g_{03},g_{05},g_{13},g_{15},g_{23},g_{25})^T.
\end{aligned}
\label{eq:spinful_coupling_order}
\end{equation}
The PH source rows use the orbital order
\begin{equation}
\mathcal B_{ph}=(\gamma_0,\gamma_1,\gamma_2,\gamma_3,\gamma_5,
\gamma_{01},\gamma_{02},\gamma_{03},\gamma_{05},\gamma_{12},
\gamma_{13},\gamma_{15},\gamma_{23},\gamma_{25},\mathbb I,\gamma_{35}).
\label{eq:spinful_ph_source_order}
\end{equation}
The first 16 rows of $M^{ph}$ are the charge multiplets $\mathcal B_{ph}\otimes s_0$, and the last 16 rows are the spin-vector multiplets $\mathcal B_{ph}\otimes\vec s$ in the same orbital order. For the PP sources, the spin-singlet rows use symmetric orbital matrices and the spin-triplet rows use antisymmetric orbital matrices, ordered as
\begin{align}
\mathcal B_{pp}^{S}&=(\gamma_0,\gamma_2,\gamma_5,\gamma_{01},\gamma_{03},
\gamma_{12},\gamma_{15},\gamma_{23},\gamma_{35},\mathbb I),\\
\mathcal B_{pp}^{T}&=(\gamma_1,\gamma_3,\gamma_{02},\gamma_{05},\gamma_{13},\gamma_{25}).
\label{eq:spinful_pp_source_order}
\end{align}
Each triplet row denotes an $SU(2)$ vector multiplet. In these bases, the logarithmic source flows decouple:
\begin{align}
\frac{d\ln\Delta_i^{ph}}{d\ln s}
&=2+\mathcal D_0\sum_{a=1}^{16}M_{ia}^{ph}g_a^c,
\quad i=1,\ldots,32,\\
\frac{d\ln\Delta_i^{pp}}{d\ln s}
&=2+\mathcal D_0\sum_{a=1}^{16}M_{ia}^{pp}g_a^c,
\quad i=1,\ldots,16.
\label{eq:spinful_source_flow}
\end{align}
Equivalently, for $X=ph,pp$,
\begin{equation}
\frac{d\boldsymbol\Delta^X}{d\ln s}
=\left[2I_{N_X}+\mathcal D_0\operatorname{Diag}(M^X\mathbf g^c)\right]
\boldsymbol\Delta^X,
\qquad N_{ph}=32,\quad N_{pp}=16.
\label{eq:spinful_source_flow_vector}
\end{equation}
Here $M^{ph}\in\mathbb R^{32\times16}$ and $M^{pp}\in\mathbb R^{16\times16}$ map the coupling vector to the one-loop correction in each source channel. The source-space operators in Eq.~\eqref{eq:spinful_source_flow_vector} are the corresponding diagonal $32\times32$ and $16\times16$ matrices. The coefficient matrices are
\begin{equation}
M^{ph}=\left(
\begin{array}{cccccccccccccccc}
-7 & -1 & -1 & 1 & -1 & -1 & 1 & 1 & -1 & -1 & -1 & -1 & 1 & 1 & 1 & 1 \\
-2 & -2 & -2 & -2 & -14 & 2 & 2 & 2 & -2 & -2 & 2 & 2 & -2 & -2 & 2 & 2 \\
-1 & 1 & -7 & -1 & -1 & -1 & 1 & 1 & -1 & -1 & 1 & 1 & 1 & 1 & -1 & -1 \\
-2 & 2 & -2 & 2 & -2 & 2 & 2 & -2 & -14 & -2 & -2 & 2 & -2 & 2 & -2 & 2 \\
-2 & 2 & -2 & 2 & -2 & 2 & 2 & -2 & -2 & -14 & 2 & -2 & 2 & -2 & 2 & -2 \\
-1 & -7 & 1 & -1 & -1 & -1 & 1 & 1 & 1 & 1 & -1 & -1 & -1 & -1 & 1 & 1 \\
-2 & -2 & -2 & -2 & 2 & -14 & 2 & 2 & 2 & 2 & -2 & -2 & 2 & 2 & -2 & -2 \\
-1 & -1 & 1 & 1 & 1 & -1 & 1 & -1 & -1 & 1 & -7 & -1 & -1 & 1 & -1 & 1 \\
-1 & -1 & 1 & 1 & 1 & -1 & 1 & -1 & 1 & -1 & -1 & -7 & 1 & -1 & 1 & -1 \\
1 & -1 & -1 & -7 & -1 & -1 & 1 & 1 & 1 & 1 & 1 & 1 & -1 & -1 & -1 & -1 \\
0 & 0 & 0 & 0 & 0 & 0 & 0 & 0 & 0 & 0 & 0 & 0 & 0 & 0 & 0 & 0 \\
0 & 0 & 0 & 0 & 0 & 0 & 0 & 0 & 0 & 0 & 0 & 0 & 0 & 0 & 0 & 0 \\
1 & 1 & -1 & -1 & 1 & -1 & 1 & -1 & -1 & 1 & -1 & 1 & -1 & 1 & -7 & -1 \\
1 & 1 & -1 & -1 & 1 & -1 & 1 & -1 & 1 & -1 & 1 & -1 & 1 & -1 & -1 & -7 \\
0 & 0 & 0 & 0 & 0 & 0 & 0 & 0 & 0 & 0 & 0 & 0 & 0 & 0 & 0 & 0 \\
0 & 0 & 0 & 0 & 0 & 0 & 0 & 0 & 0 & 0 & 0 & 0 & 0 & 0 & 0 & 0 \\
1 & -1 & -1 & 1 & -1 & -1 & 1 & 1 & -1 & -1 & -1 & -1 & 1 & 1 & 1 & 1 \\
-2 & -2 & -2 & -2 & 2 & 2 & 2 & 2 & -2 & -2 & 2 & 2 & -2 & -2 & 2 & 2 \\
-1 & 1 & 1 & -1 & -1 & -1 & 1 & 1 & -1 & -1 & 1 & 1 & 1 & 1 & -1 & -1 \\
-2 & 2 & -2 & 2 & -2 & 2 & 2 & -2 & 2 & -2 & -2 & 2 & -2 & 2 & -2 & 2 \\
-2 & 2 & -2 & 2 & -2 & 2 & 2 & -2 & -2 & 2 & 2 & -2 & 2 & -2 & 2 & -2 \\
-1 & 1 & 1 & -1 & -1 & -1 & 1 & 1 & 1 & 1 & -1 & -1 & -1 & -1 & 1 & 1 \\
-2 & -2 & -2 & -2 & 2 & 2 & 2 & 2 & 2 & 2 & -2 & -2 & 2 & 2 & -2 & -2 \\
-1 & -1 & 1 & 1 & 1 & -1 & 1 & -1 & -1 & 1 & 1 & -1 & -1 & 1 & -1 & 1 \\
-1 & -1 & 1 & 1 & 1 & -1 & 1 & -1 & 1 & -1 & -1 & 1 & 1 & -1 & 1 & -1 \\
1 & -1 & -1 & 1 & -1 & -1 & 1 & 1 & 1 & 1 & 1 & 1 & -1 & -1 & -1 & -1 \\
0 & 0 & 0 & 0 & 0 & 0 & 0 & 0 & 0 & 0 & 0 & 0 & 0 & 0 & 0 & 0 \\
0 & 0 & 0 & 0 & 0 & 0 & 0 & 0 & 0 & 0 & 0 & 0 & 0 & 0 & 0 & 0 \\
1 & 1 & -1 & -1 & 1 & -1 & 1 & -1 & -1 & 1 & -1 & 1 & -1 & 1 & 1 & -1 \\
1 & 1 & -1 & -1 & 1 & -1 & 1 & -1 & 1 & -1 & 1 & -1 & 1 & -1 & -1 & 1 \\
0 & 0 & 0 & 0 & 0 & 0 & 0 & 0 & 0 & 0 & 0 & 0 & 0 & 0 & 0 & 0 \\
0 & 0 & 0 & 0 & 0 & 0 & 0 & 0 & 0 & 0 & 0 & 0 & 0 & 0 & 0 & 0 \\
\end{array}
\right)
\end{equation}

\begin{equation}
M^{pp}=\left(
\begin{array}{cccccccccccccccc}
-1 & 1 & 1 & -1 & -1 & -1 & -1 & -1 & -1 & 1 & 1 & -1 & 1 & -1 & -1 & 1 \\
1 & -1 & -1 & 1 & -1 & -1 & -1 & -1 & -1 & 1 & -1 & 1 & 1 & -1 & 1 & -1 \\
0 & 0 & 0 & 0 & 0 & 0 & 0 & 0 & 0 & 0 & 0 & 0 & 0 & 0 & 0 & 0 \\
1 & -1 & -1 & 1 & -1 & -1 & -1 & -1 & 1 & -1 & 1 & -1 & -1 & 1 & -1 & 1 \\
1 & 1 & -1 & -1 & 1 & -1 & -1 & 1 & -1 & -1 & -1 & -1 & -1 & -1 & 1 & 1 \\
-1 & 1 & 1 & -1 & -1 & -1 & -1 & -1 & 1 & -1 & -1 & 1 & -1 & 1 & 1 & -1 \\
-2 & 2 & -2 & 2 & -2 & 2 & -2 & 2 & 2 & 2 & -2 & -2 & -2 & -2 & -2 & -2 \\
-1 & -1 & 1 & 1 & 1 & -1 & -1 & 1 & -1 & -1 & 1 & 1 & -1 & -1 & -1 & -1 \\
-2 & -2 & -2 & -2 & 2 & 2 & -2 & -2 & -2 & 2 & 2 & -2 & -2 & 2 & 2 & -2 \\
-2 & -2 & -2 & -2 & 2 & 2 & -2 & -2 & 2 & -2 & -2 & 2 & 2 & -2 & -2 & 2 \\
0 & 0 & 0 & 0 & 0 & 0 & 0 & 0 & 0 & 0 & 0 & 0 & 0 & 0 & 0 & 0 \\
0 & 0 & 0 & 0 & 0 & 0 & 0 & 0 & 0 & 0 & 0 & 0 & 0 & 0 & 0 & 0 \\
0 & 0 & 0 & 0 & 0 & 0 & 0 & 0 & 0 & 0 & 0 & 0 & 0 & 0 & 0 & 0 \\
1 & 1 & -1 & -1 & 1 & -1 & -1 & 1 & 1 & 1 & 1 & 1 & 1 & 1 & -1 & -1 \\
-2 & 2 & -2 & 2 & -2 & 2 & -2 & 2 & -2 & -2 & 2 & 2 & 2 & 2 & 2 & 2 \\
-1 & -1 & 1 & 1 & 1 & -1 & -1 & 1 & 1 & 1 & -1 & -1 & 1 & 1 & 1 & 1 \\
\end{array}
\right)
\end{equation}

\section{Bethe--Salpeter analysis at finite doping}
\label{sec:app:finite_doping_bse}

At finite doping, the $\Gamma$ and $M$ QBTs develop Fermi pockets that control
the available particle-particle (PP) and particle-hole (PH) phase space.  We
solve a full-component static BSE using exact band energies and Bloch
eigenvectors in a common low-energy window.  The three density-density interactions are held fixed as $\mu$ changes.  Here $\QQ=0,\mathbf M$ denotes the pair momentum in a PP block and the
transfer momentum in a PH block.

\subsection{Nonsymmorphic sewing and the bare Cooper response}

For $v=\Gamma,M$, let $\mathbf K_\Gamma=0$ and
$\mathbf K_M=\mathbf M$.  The retained exact-band energies measured from the
chemical potential are
\begin{equation}
    \xi_{v,n}(\pp;\mu)
    =E_{v,n}(\mathbf K_v+\pp)-\mu,
    \qquad
    \mathbf K_\Gamma=0,
    \quad
    \mathbf K_M=\mathbf M,
    \label{eq:physical_bse_xi}
\end{equation}
where $n$ labels the symmetry-related bands in the two valleys.  In these
coordinates, an intervalley pair has total momentum $\QQ=\mathbf M$,
\begin{equation}
    \Delta_M^\dagger[\Phi]
    =\sum_{\pp,\alpha\beta}
    \Phi_{\alpha\beta}(\pp)
    c^\dagger_{\Gamma,\pp,\alpha}
    c^\dagger_{M,-\pp,\beta}.
    \label{eq:physical_bse_pdw_source}
\end{equation}
The nonsymmorphic sewing relation in Eq.~\eqref{eq:app_ns_pdw_nesting} enforces
equal energies for the two states in this pair.  The thermal factors entering
the bare PP and PH susceptibilities are
\begin{align}
    L_i^{PP}
    &=\frac{1-f(\xi_{\nu,\kk})-f(\xi_{\nu',\QQ-\kk})}
    {\xi_{\nu,\kk}+\xi_{\nu',\QQ-\kk}},\label{eq:physical_bse_loop_factors_pp}
    \\
    L_i^{PH}
    &=\frac{f(\xi_{\nu,\kk})-f(\xi_{\nu',\kk+\QQ})}
    {\xi_{\nu',\kk+\QQ}-\xi_{\nu,\kk}}.
    \label{eq:physical_bse_loop_factors_ph}
\end{align}
For a symmetry-matched $\Gamma$--$M$ pair at $\QQ=\mathbf M$,
$\xi_{\nu,\kk}=\xi_{\nu',\QQ-\kk}\equiv\xi$ along the Fermi contour.  The
PP factor therefore becomes $\tanh(\xi/2T)/(2\xi)$, and its radial integral
produces the usual Cooper logarithm.
The bare $\QQ=\mathbf M$ intervalley Cooper susceptibility therefore has the
same logarithmic divergence as its intravalley $\QQ=0$ counterpart, and this is protected by the nonsymmorphic symmetry.  The interaction and Bloch form factors then determine the leading PP eigenmode.

\subsection{Local interaction and exact-band projection}

The fixed interaction in the valley--QBT-spinor convention is
\begin{equation}
    S_{\rm int}
    =\frac12\sum_{\ell=I,13,15}g_\ell
    \int d\tau d^2\mathbf r\,
    \left(\psi^\dagger\Gamma_\ell s_0\psi\right)^2,
    \label{eq:physical_bse_interaction}
\end{equation}
where $s_0$ is omitted for spinless fermions and
$\Gamma_I=\tau_0\sigma_0$, $\Gamma_{13}=\tau_y\sigma_0$, and
$\Gamma_{15}=\tau_x\sigma_0$.  We write
$U_\ell=g_\ell/\mathcal D_{\rm ref}$ with
$\mathcal D_{\rm ref}=1/(8\pi t_{\rm ref})=1/(16\pi t_0)$ and
$t_{\rm ref}\equiv t_{\rm mic}=2t_0$.  Here $\mathcal D_{\rm ref}$ fixes the
common normalization of the vertex formed by these three couplings.  This choice rescales all BSE eigenvalues uniformly and does not change their ordering.  Antisymmetrizing the two outgoing fermion legs gives the common vertex
\begin{equation}
    V_{ab;cd}
    =\sum_\ell U_\ell
    \left[
    (\Gamma_\ell)_{ac}(\Gamma_\ell)_{bd}
    -(\Gamma_\ell)_{ad}(\Gamma_\ell)_{bc}
    \right].
    \label{eq:physical_bse_vertex}
\end{equation}

We label a physical block by $c=(\eta,\QQ,\mathcal S)$, where $\eta=PP,PH$ specifies the channel, $\QQ=0,\mathbf M$ its pair or transfer momentum, and $\mathcal S$ the applicable spin block: spinless, singlet/triplet for PP, or charge/spin for PH.  Because the \(\Gamma\) and \(M\) valleys acquire opposite phases under a primitive moiré translation, the \(T_{\mathbf R}\) parity in Table~\ref{tab:symmetry_classification} fixes the momentum carried by a local matrix: translation-even components belong to the \(\QQ=0\) sector, whereas translation-odd components belong to the \(\QQ=\mathbf M\) sector. Within each such sector, we retain all matrix components compatible with the specified spin block and, for PP, with Fermi antisymmetry, allowing them to mix in the BSE. In the uniform PH charge sector, the identity component represents the conserved total density rather than a symmetry-breaking order and is therefore excluded from the instability comparison and evaluated separately as the compressibility. The momentum-independent matrices $F_A^{(c)}$ act in
the local valley--QBT-spinor space.  In a PP block they specify a pair with
total momentum $\QQ$ and the appropriate spin factor; in a PH block they
specify a density or excitonic bilinear with transfer momentum $\QQ$.  We use
the orthonormal basis
\begin{equation}
    \Tr\!\left[F_A^{(c)\dagger}F_B^{(c)}\right]=\delta_{AB}.
    \label{eq:physical_bse_hs_basis}
\end{equation}
Then, the antisymmetrized vertex $V_{ab;cd}$ is used to construct both the PP and PH kernels. Its action on a local basis matrix \(F\) is
\begin{align}
 \mathcal A^{PP}[F]_{\alpha\beta}
    &=-\frac12 V_{\alpha\beta;\gamma\delta}F_{\gamma\delta},
    \\
    \mathcal A^{PH}[F]_{\alpha\gamma}
    &=-V_{\alpha\beta;\gamma\delta}F_{\delta\beta},
    \label{eq:physical_bse_actions}
\end{align}
where \(\alpha,\beta,\gamma,\delta\) include spin when present. Expanding the local order matrices in the orthonormal basis \(\{F_A^{(c)}\}\), the corresponding channel interaction operator is represented in coefficient space by
\begin{equation*}
    A_{AB}^{(c)}
    =\Tr\!\left[
    F_A^{(c)\dagger}\mathcal A^{(c)}[F_B^{(c)}]
    \right].
\end{equation*}

In a PP block, the pairing matrix obeys
\begin{equation}
    \Phi_\QQ(\kk)=-\Phi_\QQ^T(\QQ-\kk).
    \label{eq:physical_bse_fermion_antisymmetry}
\end{equation}
Accordingly, the orbital pairing space contains six components for spinless fermions, ten for spin-singlet pairing, and six for spin-triplet pairing, consistent with the RG classification. These components are further partitioned between the \(\QQ=0\) and \(\QQ=\mathbf M\) sectors by their translation parity described above. In the spin-triplet sector, global \(SU(2)\) symmetry makes the three spin orientations degenerate, so we diagonalize one representative component and assign the resulting mode a threefold spin multiplicity.

No analogous antisymmetry constraint applies to a PH bilinear. Each charge or spin sector therefore contains 16 orbital matrices before they are partitioned by translation parity. For \(\QQ=\mathbf M\), the two intervalley orientations, \(\Gamma\!\to\! M\) and \(M\!\to\!\Gamma\), are components of the same valley-off-diagonal block rather than independent channels. They are therefore included once within the same matrix basis, with no additional factor of two in the interaction kernel.

To evaluate the BSE with the exact Bloch states, we first embed each local
order matrix $F_A^{(c)}$ into the microscopic layer--sublattice plane-wave
basis, obtaining $\widehat F_A^{(c)}$.  For fixed $c$, let
$i=(\nu,\nu',\kk)$ label an exact-band pair or transition.  The corresponding
matrix elements between exact Bloch states are
\begin{align}
    X_{iA}^{PP}
    &=u_{\nu,\kk}^\dagger
      \widehat F_A^{PP}
      u_{\nu',\QQ-\kk}^*,
    \\
    X_{iA}^{PH}
    &=u_{\nu',\kk+\QQ}^\dagger
      \widehat F_A^{PH}
      u_{\nu,\kk},
    \label{eq:physical_bse_transition_amplitudes}
\end{align}
where $u_{\nu,\kk}$ denotes the exact Bloch eigenvector in the microscopic
basis.  The embedding fixes the local order structure inherited from the QBT
theory, while the exact Bloch eigenvectors generate its momentum-dependent
form factor.

Combining these projected amplitudes with the corresponding PP or PH loop
factors in Eq.~\eqref{eq:physical_bse_loop_factors_pp} and Eq.~\eqref{eq:physical_bse_loop_factors_ph} give the bare
susceptibility in the local order basis,
\begin{equation}
    \omega_i^{(c)}
    =\frac{d^2k_i}{(2\pi)^2}L_i^{(c)},
    \qquad
    \Pi_{AB}^{(c)}
    =\sum_i
    \omega_i^{(c)}
    X_{iA}^{(c)*}X_{iB}^{(c)}.
    \label{eq:physical_bse_bubble_matrix}
\end{equation}
Thus $\Pi_{AB}^{(c)}$ measures the bare response between the local order
components $F_A^{(c)}$ and $F_B^{(c)}$ after their projection onto the exact
bands.

For a general local order matrix
$F=\sum_A f_A F_A^{(c)}$, the coefficient vector $f$ therefore has bare
static susceptibility $f^\dagger\Pi_c f$, while $\Pi_c f$ gives the
corresponding induced polarization in the same local-order coefficient space.

\subsection{Hermitian BSE and numerical implementation}

Since the loop weights are nonnegative, the susceptibility matrix constructed
above is positive semidefinite, $\Pi_c\succeq0$.  Components in the null space
of $\Pi_c$ have no bare response and can therefore be removed from the BSE.  Let $P_c$ denote the projector onto $\operatorname{supp}\Pi_c$, and define the interaction restricted to this support by
\begin{equation*}
    \bar A^{(c)}=P_c A^{(c)} P_c.
\end{equation*}
For notational simplicity, we henceforth drop the bar.  In the same
local-order coefficient basis $\{F_A^{(c)}\}$ used above, the BSE then takes
the form
\begin{equation}
    A^{(c)}\Pi_c f_{c,n}
    =\lambda_{c,n}f_{c,n}.
    \label{eq:physical_bse_local_eigenproblem}
\end{equation}

Although both $A^{(c)}$ and $\Pi_c$ are Hermitian, their product $A^{(c)}\Pi_c$ is
not necessarily Hermitian. On $\operatorname{supp}\Pi_c$, the eigenproblem can instead be written in the equivalent Hermitian form
\begin{equation}
    K_c=\Pi_c^{1/2}A^{(c)}\Pi_c^{1/2},
    \qquad
    K_c y_{c,n}=\lambda_{c,n}y_{c,n},
    \qquad
    y_{c,n}=\Pi_c^{1/2}f_{c,n},
    \qquad
    f_{c,n}=\Pi_c^{-1/2}y_{c,n},
    \label{eq:physical_bse_kernel}
\end{equation}
as in the symmetrized form of standard linearized gap equations
\cite{PhysRevB.98.174503}. The two eigenproblems have the same nonzero eigenvalues and corresponding modes.  Since $A^{(c)}$ is Hermitian, $K_c$ is Hermitian and all $\lambda_{c,n}$ are real.

We diagonalize $K_c$ independently in each physical block
$c=(\eta,\QQ,\mathcal S)$. The largest algebraic eigenvalue $\lambda_c$ defines the
leading mode of the block.  Its coefficient vector determines the corresponding local order matrix,
\begin{equation}
    \Phi_{c,n}^{\rm loc}
    =\sum_A f_{c,n,A}F_A^{(c)}.
    \label{eq:physical_bse_order_profile}
\end{equation}
The physical block $c$ specifies the PP/PH, momentum, and spin sector, while
the dominant component of $\Phi_{c,n}^{\rm loc}$ is used to label the order.
Its momentum-dependent profile in the exact-band basis is encoded by the
projected amplitudes $X_{iA}^{(c)}$ defined above.

The PP and PH kernels are evaluated within the same low-energy exact-band
window,
\begin{equation}
    |E_{v,n}(\mathbf K_v+\pp)|\leq E_c ,
\end{equation}
where $E$ is measured relative to the QBT energy, while
$\xi=E-\mu$ in Eq.~\eqref{eq:physical_bse_xi} is measured relative to the
chemical potential.  A pair or particle--hole transition is included only
when both constituent states lie within this window.  For the parameters used
here, the window contains only the relevant low-energy conduction and valence
bands and fully covers the Fermi pockets considered.

We use $T=0.001$ and $E_c=0.1$ as the reference parameters.
The BSE spectra are numerically converged with respect to plane-wave
truncation, retained band space, and momentum quadrature.
Representative-point checks at $T=0.0005,0.001,0.002$ with $E_c=0.1$,
and at $E_c=0.08,0.10,0.12$ with $T=0.001$, preserve the leading-family
assignments away from unresolved boundaries.

\subsection{Resolution criteria and finite-doping results}

Having obtained the leading eigenvalue and eigenvector in each physical block,
we next specify how the competing BSE modes are compared and assigned in the
phase diagrams.  Each block is treated as a distinct order family, except for
an exact charge--spin degeneracy in the spinful PH sector.  For a given local
orbital matrix $F$, the corresponding charge and spin bilinears are
$\psi^\dagger F s_0\psi$ and $\psi^\dagger F s_\alpha\psi$
($\alpha=x,y,z$), respectively.  If the charge and spin blocks have the same
$\QQ$, the same leading orbital matrix $F$, and exactly degenerate leading
eigenvalues for the retained interaction, we group them into a single family
and assign a joint label, such as QAH/QSH.  Otherwise, they are kept as
distinct competing families.

Let $\Lambda_{\mathcal F}$ denote the leading BSE eigenvalue of a family
$\mathcal F$: it is $\lambda_{c,1}$ for a single block and the common
eigenvalue for an exactly degenerate charge--spin pair.  After this grouping,
let $\Lambda_{(1)}$ and $\Lambda_{(2)}$ be the two largest eigenvalues among
inequivalent families.  To avoid assigning a unique leading instability when
two families are nearly degenerate or numerically unresolved, we require
\begin{equation}
    \Lambda_{(1)}-\Lambda_{(2)}>
    \max\!\left[
    0.05\max(|\Lambda_{(1)}|,|\Lambda_{(2)}|),
    \delta\Lambda_{(1)}+\delta\Lambda_{(2)}
    \right].
    \label{eq:physical_bse_separation}
\end{equation}
Thus the leading family must exceed its nearest competitor by both a $5\%$
relative separation and the combined discretization uncertainty.  Points that
do not satisfy this criterion are left unresolved and shown in gray.  Exact
charge--spin degeneracies are grouped before this comparison and therefore do
not by themselves generate gray points.

Once a leading family is resolved, we identify its internal order from the
dominant component of the corresponding local coefficient vector.  Defining
\begin{equation*}
    p_A=
    \frac{|f_{c,1,A}|^2}
         {\sum_B |f_{c,1,B}|^2},
\end{equation*}
the order label is assigned by the component with the largest weight $p_A$.
In all resolved regions this dominant component is well separated from the
subleading ones, so the resulting order labels are unambiguous.

This component-wise dominance is distinct from spectral isolation within the
same physical block.  We therefore separately compare the two largest
eigenvalues of the block.  Points satisfying
\begin{equation*}
    \lambda_{c,1}-\lambda_{c,2}\leq 2\delta K_c
\end{equation*}
are marked by a dotted overlay.  Here $\delta K_c$, as well as the
inter-family uncertainty $\delta\Lambda$ introduced above, is estimated by
comparing independent polar and Cartesian momentum quadratures.  Such points
still have a resolved leading family, but the detailed local order profile may
rotate within a nearly degenerate eigenspace.

\begin{figure}[htbp]
\centering
\includegraphics[width=\textwidth,keepaspectratio]
{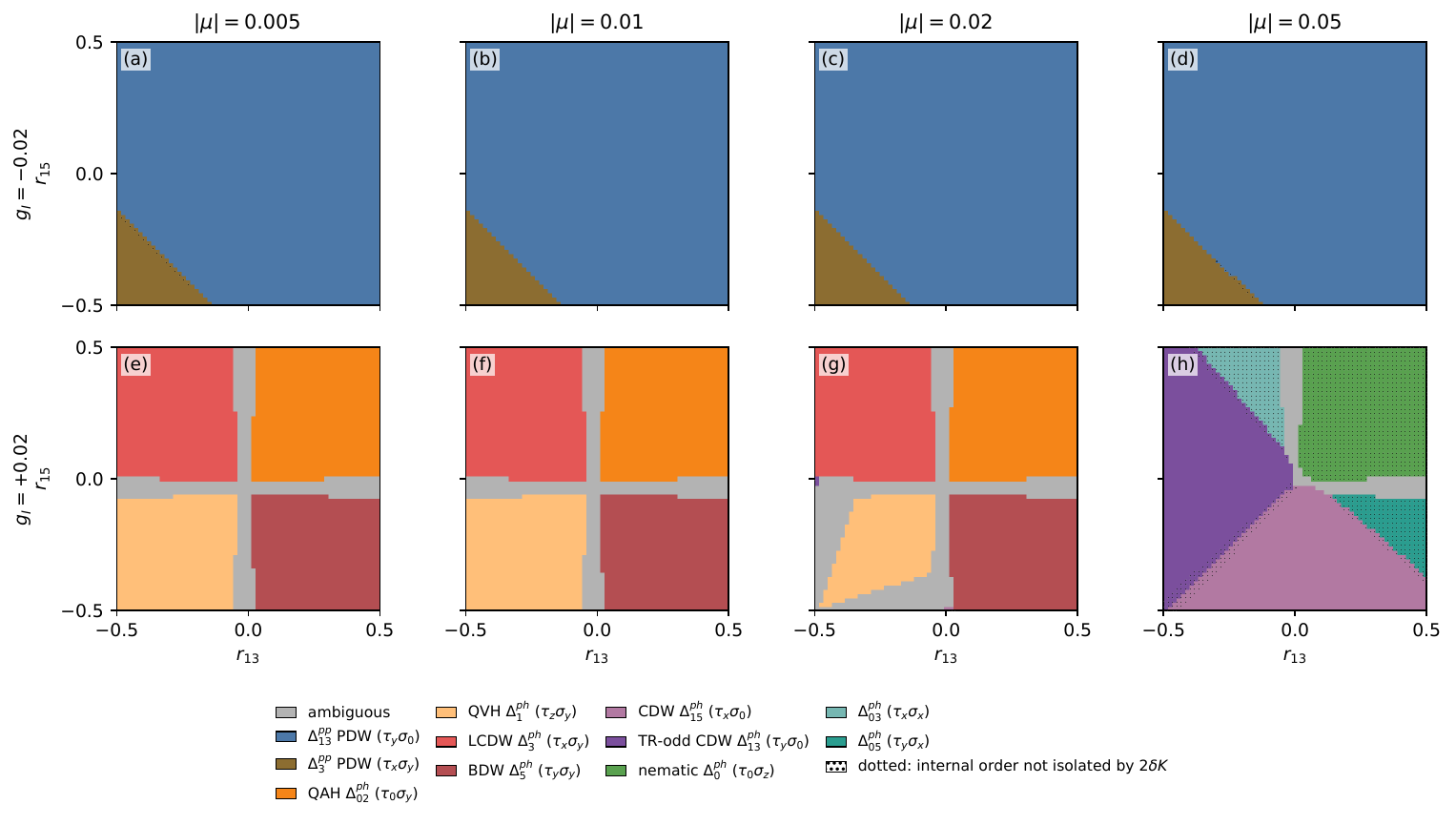}
\caption{Spinless full-component BSE at finite doping for
$g_I=-0.02,+0.02$ (rows) and
$|\mu|=0.005,0.010,0.020,$ and $0.050$ (columns),
with $T=0.001$ and $E_c=0.1$.  The axes are $r_{13}=g_{13}/|g_I|$ and
$r_{15}=g_{15}/|g_I|$.  Each panel compares the antisymmetric PP sectors at
$\QQ=0,\mathbf M$ with PH sectors at $\QQ=0,\mathbf M$ using
Eq.~\eqref{eq:physical_bse_kernel}.  Colors identify the resolved leading
family and its dominant $\tau_i\sigma_j$ component using the order names of the
RG classification.
The low-doping interband response lies in the $\sigma_y$ mass sector, while
the higher-doping map also contains $\sigma_0$, $\sigma_x$, and $\sigma_z$
branches.  Gray points do not satisfy
Eq.~\eqref{eq:physical_bse_separation}; the dotted overlay instead marks a
resolved sector whose internal order matrix is not isolated by the pointwise
$2\delta K_c$ criterion.}
\label{fig:finite_doping_bse_spinless}
\end{figure}

Fig.~\ref{fig:finite_doping_bse_spinless} summarizes the spinless
finite-doping results, while the spinful phase diagram is shown in
Fig.~\ref{fig:main_finite_doping_bse_spinful}.  For $g_I<0$, the leading
response remains predominantly in the PP sector, continuously connecting to
the Cooper tendency found in the charge-neutral RG.  Nonsymmorphic sewing
gives the $\QQ=0$ and $\QQ=\mathbf M$ Cooper channels the same logarithmic
enhancement, while the interaction vertex and exact-band Bloch form factors
select the antisymmetric $\Delta_{13}^{pp}$ PDW with internal matrix
$\tau_y\sigma_0$ over most of the resolved spinless region.

For $g_I>0$, doping reorganizes the PH response by shifting spectral weight
from interband QBT transitions to intraband Fermi-pocket excitations.  At
$\QQ=0$, the two contributions can be separated explicitly as
\begin{align}
    \Pi_{AB}^{PH}(0)
    ={}&\sum_\nu\int\frac{d^2k}{(2\pi)^2}
    \bigl[-f'(\xi_{\nu,\kk})\bigr]
    X_{\nu\nu,A}^{PH*}(\kk)X_{\nu\nu,B}^{PH}(\kk)
    \nonumber\\
    &+\sum_{\nu\ne\nu'}\int\frac{d^2k}{(2\pi)^2}
    \frac{f(\xi_{\nu,\kk})-f(\xi_{\nu',\kk})}
    {\xi_{\nu',\kk}-\xi_{\nu,\kk}}
    X_{\nu\nu',A}^{PH*}(\kk)X_{\nu\nu',B}^{PH}(\kk).
    \label{eq:physical_bse_q0_ph_decomposition}
\end{align}
Near charge neutrality, the interband term dominates and favors the
$\sigma_y$ mass manifold, giving the QAH/QVH orders at $\QQ=0$ and the
LCDW/BDW orders at $\QQ=\mathbf M$.  As the Fermi pockets grow, the intraband
term becomes increasingly important and favors density-like $\sigma_0$
form factors, driving the response toward the $\QQ=\mathbf M$ CDW/SDW
sector.  The same redistribution occurs in the intervalley
$\QQ=\mathbf M$ block and accounts for the crossover seen in the finite-doping
maps.  Thus the spinless PH response evolves from the QBT mass competition
inherited from charge neutrality to a pocket-driven density-wave regime.

In the spinful PH blocks, the exchange contribution is identical in the
charge and spin channels, while the charge channel also contains the direct
contraction.  For the $\sigma_y$ masses,
\begin{equation}
    \Tr(\Gamma_\ell F)=0,
    \qquad \ell=I,13,15.
    \label{eq:physical_bse_charge_spin_degeneracy}
\end{equation}
Because each $\Gamma_\ell$ is proportional to $\sigma_0$, their charge and
spin partners are exactly degenerate.  By contrast, $\tau_x\sigma_0$ and
$\tau_y\sigma_0$ overlap with $\Gamma_{15}$ and $\Gamma_{13}$, respectively,
allowing the direct term to split the density-like CDW and SDW branches.

